\documentclass[aip,rsi,reprint,nofootinbib]{revtex4-1}
\usepackage{graphicx}
\usepackage{caption}
\usepackage{subcaption}
\usepackage[]{graphicx}
\usepackage[]{color}
\usepackage[]{mathtools}
\usepackage{enumerate}
\usepackage{float}
\usepackage{amsmath}
\usepackage{amssymb}
\usepackage{amsfonts}
\usepackage{mathrsfs}
\usepackage[unicode=true,colorlinks=no,pdfborder=no]{hyperref}  %
\usepackage{makeidx}
\usepackage{latexsym}
\usepackage{color}
\usepackage{multirow}
\usepackage{float}
\usepackage{dcolumn}
\usepackage{bm}
\usepackage[mathlines]{lineno}
\begin{document}

\title{Readout for Kinetic-Inductance-Detector-Based Submillimeter Radio Astronomy}

\author{Ran Duan, Xinxin Zhang, Chenhui Niu and Di Li}
\email[]{duanran@nao.cas.cn}

\affiliation{National Astronomical Observatories, Chinses Academy of Science, Beijing 100012, China}

\date{\today}

\begin{abstract}

A substantial amount of important scientific information is contained within astronomical data at the submillimeter and far-infrared (FIR) wavelengths, including information regarding dusty galaxies, galaxy clusters, and star-forming regions; however, these wavelengths are among the least-explored fields in astronomy because of the technological difficulties involved in such research. Over the past 20 years, considerable efforts have been devoted to developing submillimeter- and millimeter-wavelength astronomical instruments and telescopes.

   The number of detectors is an important property of such instruments and is the subject of the current study. Future telescopes will require as many as hundreds of thousands of detectors to meet the necessary requirements in terms of the field of view, scan speed, and resolution. A large pixel count is one benefit of the development of multiplexable detectors that use kinetic inductance detector (KID) technology.

   This paper presents the development of all aspects of the readout electronics for a KID-based instrument, which enabled one of the largest detector counts achieved to date in submillimeter-/millimeter-wavelength imaging arrays: a total of 2304 detectors. The work presented in this paper had been implemented in the MUltiwavelength Submillimeter Inductance Camera (MUSIC), a instrument for the Caltech Submillimeter Observatory (CSO) during 2013 adn 2015.

\end{abstract}






\maketitle

\section{Introduction}

\subsection{Scientific motivation}

The results of the Cosmic Background Explorer (COBE) indicate that 50\% of the luminosity and 98\% of the photons emitted by the Big Bang fall within the submillimeter and far-infrared (FIR) ranges, as shown in Fig. 1. The National Research Council's Decadal Survey (New Worlds, New Horizons in Astronomy and Astrophysics 2010) emphasizes that ``a high priority in the coming decade will be to undertake large and detailed surveys of galaxies as they evolve across the wide interval of cosmic time'' (pp. 218) and to develop, in effect, ``a 13-billion year-long movie that traces the build-up of structure since the universe first became transparent to light'' (pp. 2--13).

Considerable effort has been devoted to studying this topic over the years, including ground-based, balloon-borne, airborne, and space observatory measurements. Detection technology has also rapidly developed for constructing and upgrading instruments. The MUSIC instrument that we designed and constructed (Fig. 2 shows the major components of the instrument, which will be discussed in detail in the following sections) is ideally suited to address the priorities identified in the report and may serve as a guide for future submillimeter- and millimeter-wavelength telescopes such as the Cerro Chajnantor Atacama telescope (CCAT).

  \begin{figure}[H]
  \includegraphics[width=8.5cm]{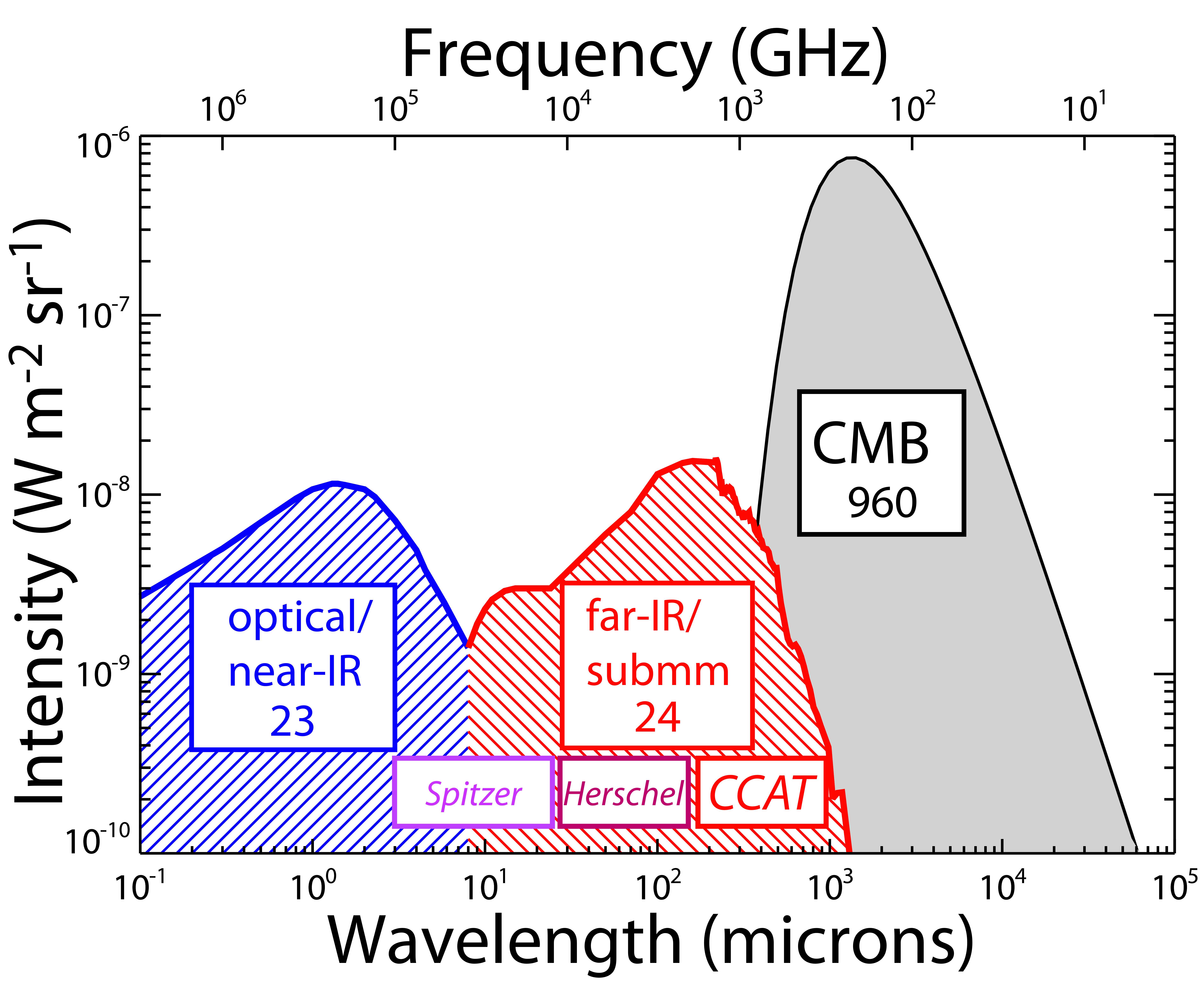}
  \caption[Integrated brightness in cosmic optical, infrared, and microwave background radiation.]
  { \label{fig:}
Integrated brightness in cosmic optical, infrared, and microwave background radiation. Reprinted from reference \cite{Dole}.}
  \end{figure}

   MUSIC is designed to have a high mapping speed, a high angular
   resolution, and four frequency bands probed by kinetic inductance
   detectors (KIDs). As the newest instrument in the Caltech
   Submillimeter Observatory, MUSIC is among the first microwave KID
   cameras and is the camera with the largest number of detectors sensitive to the submillimeter-wavelength range. MUSIC can be used to observe the Sunyaev--Zeldovich (SZ) effect in galaxy clusters, dusty star-forming galaxies, and dark matter halos to address fundamental questions regarding the large-scale structure of the universe and the history of star formation over cosmic time.
   
  \begin{figure}[H]
  \includegraphics[width=8.5cm]{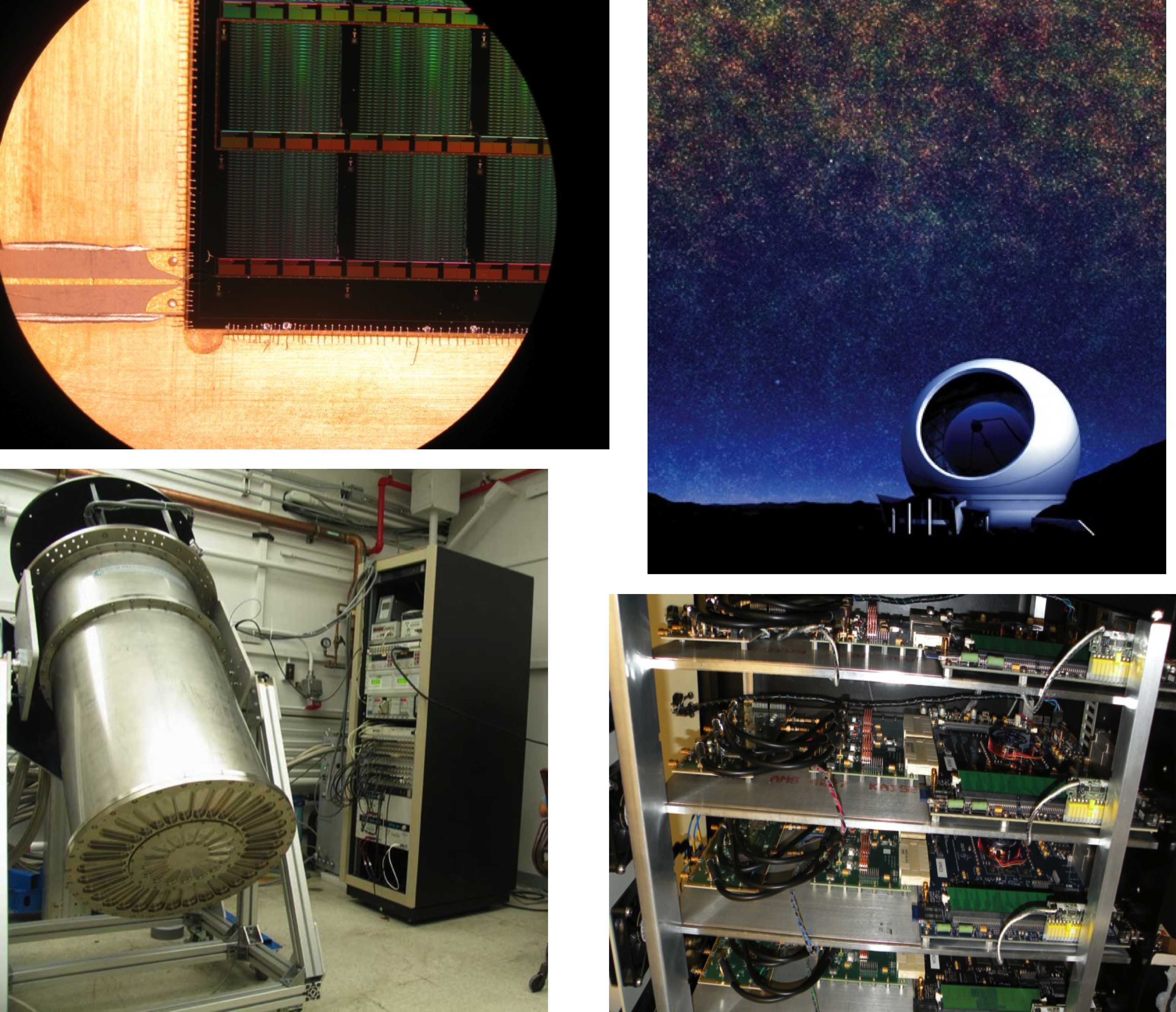}
  \caption[Photographs of various parts of the telescope instrument.]
  { \label{fig:}
Photographs of the various components of the telescope instrument. (Upper left) The detector wafer that serves as the camera of the telescope instrument. (Lower left) The cryostat that houses the detector wafer and cold electronics. (Lower right) Room-temperature readout electronics that read the signal from the detector wafer. (Upper right) The telescope in which the instrument is used for astronomical observations.}
  \end{figure}

\subsection{Submillimeter astronomy instruments}

Over the past few years, submillimeter/millimeter astronomical instruments have revolutionized our understanding of the formation of stars, galaxies, and clusters by measuring the fundamental parameters of our universe. Thus, these instruments are critical to helping us gain an understanding of nature and to continue to make new discoveries.

One commonly used detector technology for submillimeter wavelengths is the transition edge sensor (TES), which is a cryogenic sensor based on the temperature-dependent resistance of the superconducting phase transition. To read the signal from a TES, a superconducting quantum interference device (SQUID) is paired with the detector. TESs have been used for the detection of submillimeter/millimeter wavelengths in many types of instruments; however, their complex fabrication process and readout method make them difficult to scale to larger arrays. Another relatively new technology is the KID, which was developed at the Caltech/Jet Propulsion Laboratory (JPL) in the early 2000s. KIDs can be easily fabricated on a two- to three-layer wafer and frequency-domain multiplexed; all of the readout functions are performed by room-temperature electronics, with the exception of one cryogenic amplifier. KIDs are ideal for large-array implementation, which will be necessary for future telescope development; for instance, the CCAT will require millions of detectors, as shown in Fig. 3, to fully exploit its large field of view and small arcsecond resolution.

  \begin{figure}[H]
  \includegraphics[width=8.5cm]{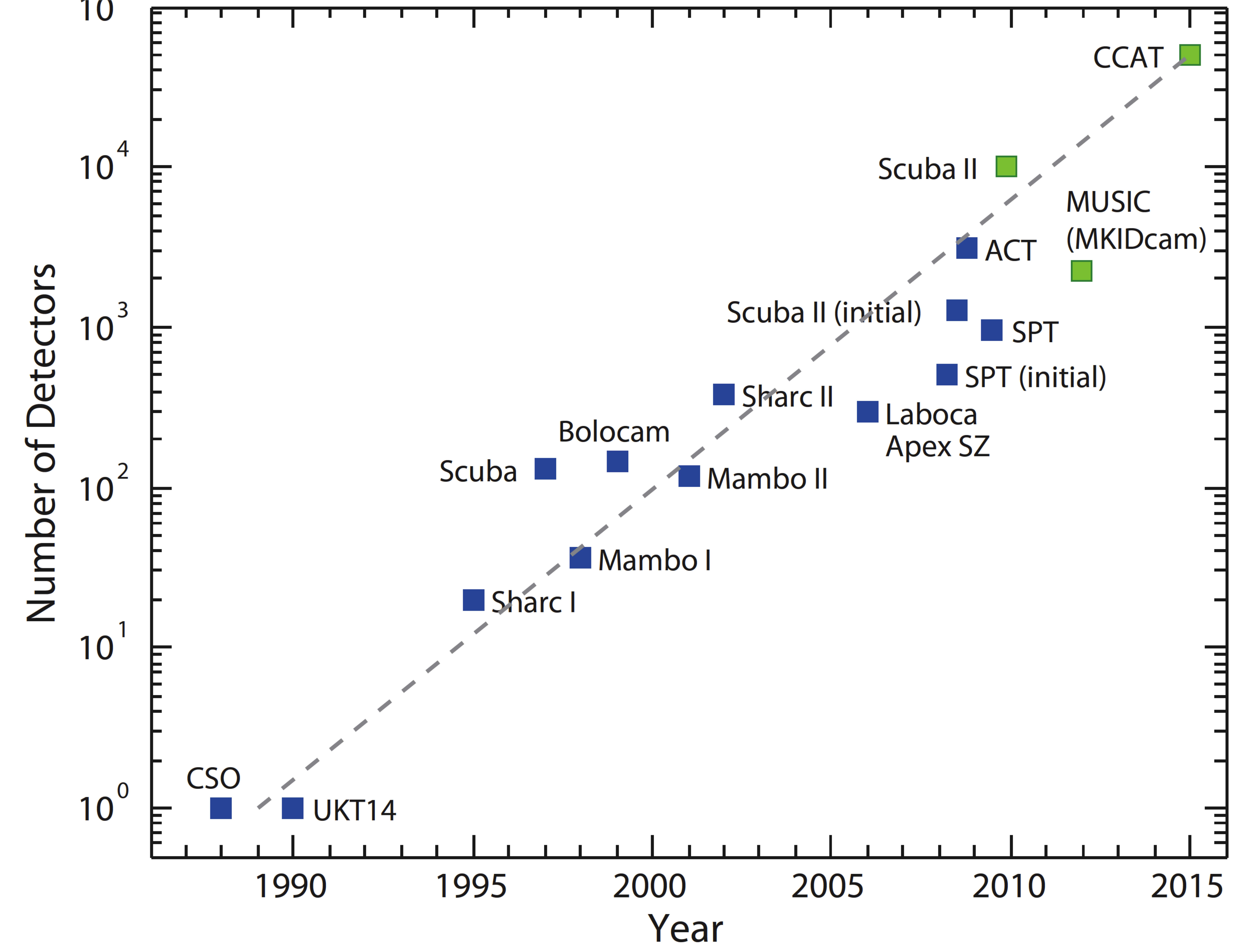}
  \caption[The growth in the size of submillimeter detector arrays as a function of time over the past two decades.]
  { \label{fig:}
The growth in the size of submillimeter detector arrays as a function of time over the past two decades. The blue points represent existing instruments, and the green points are projections at the time the plot was created. Plot courtesy of Jonas Zmuidzinas.}
  \end{figure}

The instrument that we constructed using KID technology, MUSIC, covers wavelengths of 0.87, 1.04, 1.33, and 1.98 mm. From a technological perspective, the MUSIC design successfully demonstrates a photolithographic focal plane, as illustrated in Fig. 4. This figure provides an overview of the MUSIC focal-plane wafer, which contains a broadband phased-array antenna for beam definition, a bank of four bandpass filters (BPFs) for band selection, and a microwave KID (MKID) for power detection. 

MUSIC contains the largest number of detectors of any existing KID instrument (a total of 2304 detectors) and offers large-array readout.

   MUSIC provides compelling opportunities for scientific applications including studying galaxy clusters at all redshifts at which they exist, searching for the highest-redshift contributors to the FIR background by looking for objects that are bright in the MUSIC bands but not in the SPIRE bands, and surveying protostellar cores and young stellar objects. MUSIC is also complementary to existing telescopes such as ALMA, Herschel, and Planck. It provides long-wavelength spectral-energy distribution measurements for the establishment of stricter constraints on the luminosities and dust temperatures of the high-z galaxies detected by Herschel, offering a unique probe of the high-z galaxy population and providing finding charts to the astronomical community. By providing multicolor measurements and through the combination with data from ALMA and Herschel, MUSIC survey data will stringently test models of the formation and evolution of galaxies and the dark-matter halos in which they form.

  \begin{figure}[H]
  \includegraphics[width=8.5cm]{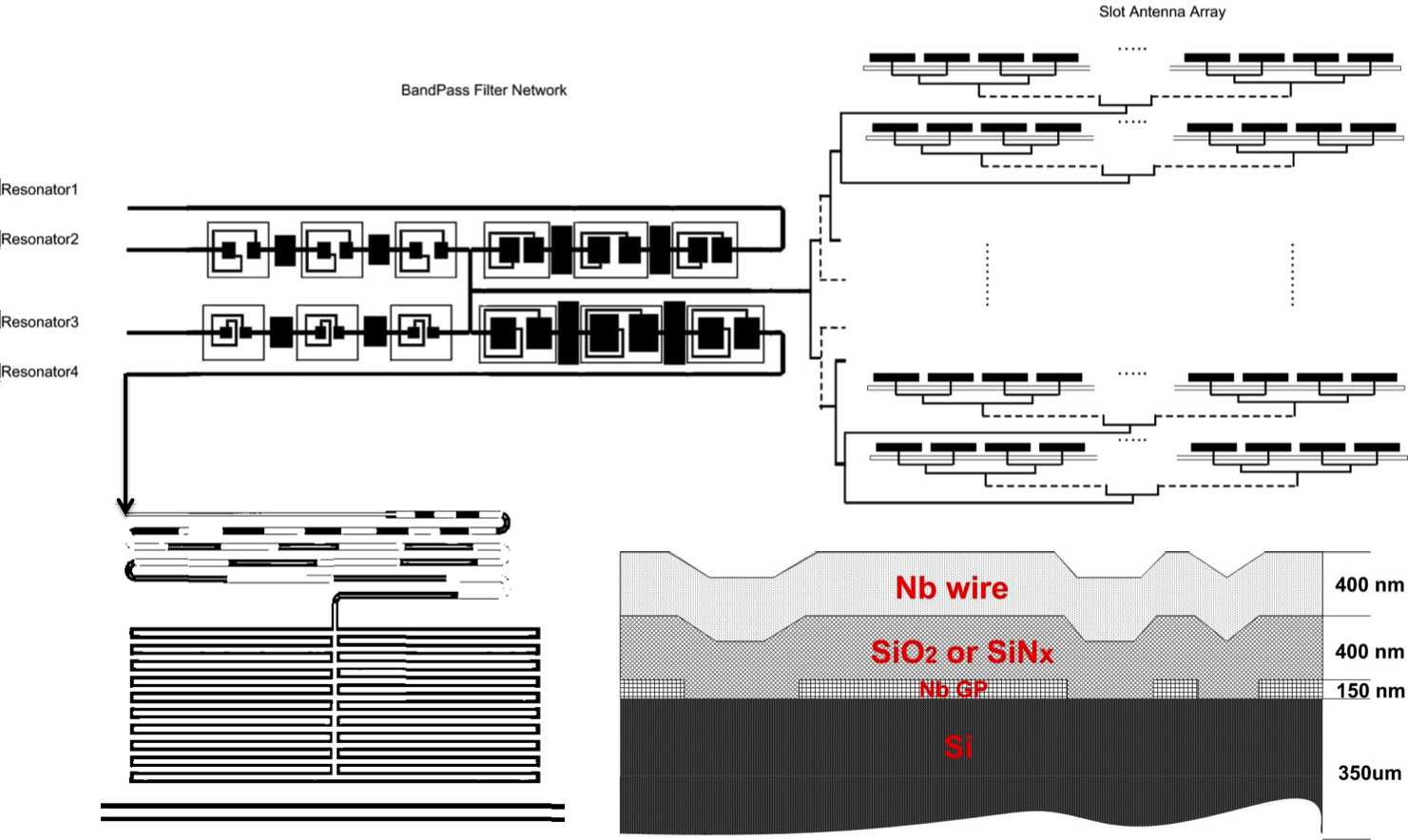}
  \caption[Scaled version of the antenna, BPF network, and KID layout.]
  { \label{fig:Scaled version of antenna, BPF network and KID layout}
 Scaled version of the antenna, BPF network, and KID layout. (Upper right) A phased-array antenna with a binary summing tree is used to capture the signal. (Upper left) Four bandpass filters split the signal from the summing tree into four different frequency bands. (Lower left) The kinetic inductance detector layout. (Lower right) Substrate layers of the detector wafer.}
  \end{figure}

\section{Readout for kinetic inductance detectors}

This section will present a systematic study of the design, implementation, and performance analysis of an open-source readout (OSR) system for a superconducting microresonator array. The OSR system performs frequency-domain multiplexed real-time complex transmission measurements to monitor the instantaneous resonance frequency and dissipation of the KIDs. With a total of 16 readout units, our OSR system can read more than 3000 complex frequency channels simultaneously. All hardware, software, and firmware were successfully installed, tested, and optimized at the Caltech Submillimeter Observatory with the first MKID camera, MUSIC, in 2010 and 2012. The system demonstrated its ability to satisfy the requirements for detector readout, data acquisition, and telescope operation. As part of the MUSIC instrument, the OSR has been in use at the CSO for scientific observation since the summer of 2013.

\subsection{Background}

The direct multiplexing of TES detectors, the other detector technology that is widely used at submillimeter and millimeter wavelengths, requires several biasing wires and one amplifier per detector. In contrast, only one input and output transmission line and one cryogenic amplifier are required for the readout of an entire KID array, thus simplifying the design of the readout circuits at the cold stage.

   Fig. 5 shows the $S_{12}$ measurements at the two ends of the transmission line for a KID array, where each dip corresponds to one resonance frequency.

  \begin{figure}[H]
  \includegraphics[width=8.5cm]{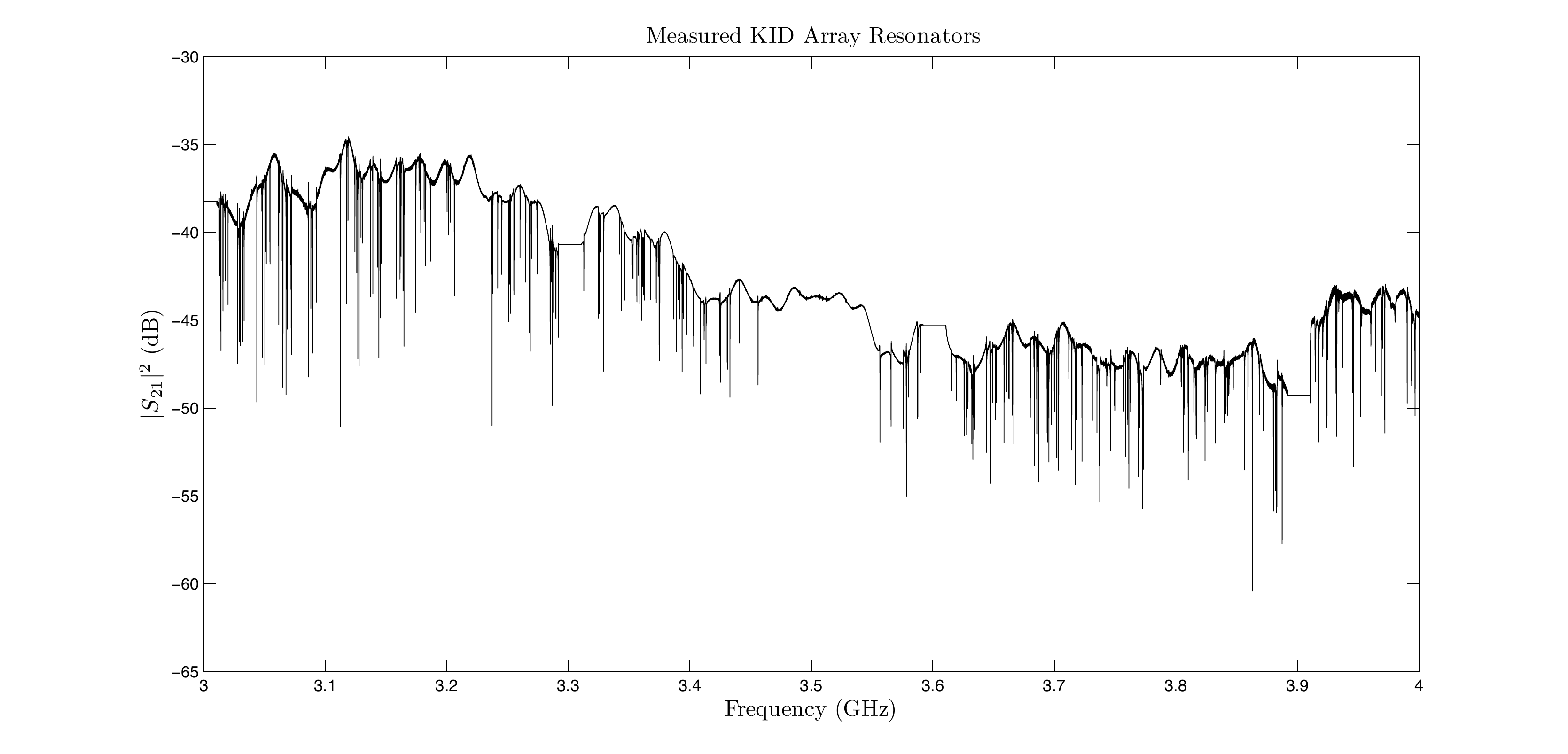}
  \caption[Multiplexed KID wafer.]
  { \label{fig:Multiplexed KID wafer}
Multiplexed KID concept: $S_{12}$ of the measured KID wafer exhibiting multiple resonance frequencies. Plot courtesy of the MUSIC group.}
  \end{figure}

      We multiplex many detectors using a single transmission line by lithographically setting slightly different resonance frequencies for each KID. The signal source (e.g., submillimeter, X-ray, FIR, ultraviolet [UV], and optical signals) modifies the microwave frequency response of the resonator. This response can be measured using a single cryogenic amplifier and sophisticated room-temperature electronics, thus shifting the complexity and challenge of the detector-array readout to the room-temperature electronics.

 Since it was first proposed at Caltech/JPL, KID technology has developed rapidly as a result of its numerous advantages and potential applications \cite{peter, James, phil, Nicole, Jack, Matt, Peter2,Mazin53}. The readout system has been demonstrated by a number of groups \cite{Mazin53,Yates53,Benz53,Irwin53,Dobbs53}. In general, the procedure for the KID OSR is as follows:
\begin{enumerate}

\item A drive tone or carrier tone (which generally ranges from approximately 100 MHz to a few gigahertz) is sent through the transmission line on the device with which the detector is coupled. The drive tone is then modulated by the detector as it responds to the power of the astronomical signal.

\item If the modulated carrier tone is at a frequency that is too high
  to be directly digitized, it is down-converted using a mixer.

\item The received tone is digitized and processed. This step includes
  digitizing the analog signal and processing the digital signal using a field-programmable gate array (FPGA), a graphics processing unit (GPU), a central processing unit (CPU), or some combination thereof. The signal from the detector generally requires real-time processing to capture the signal source, reduce the data rate, and extract useful information.

\item Auxiliary information such as a timestamp or information regarding the telescope is stored on the data acquisition (DAQ) computer.

\item Frequently, the raw data must be subjected to additional computer-based processing steps prior to use. These steps may include serialization of the data stream, noise reduction, transformation of the data format or units, calibration, or mapmaking.
\end{enumerate}

   Using the MUSIC readout as an example, this section will attempt to encompass most aspects of the readout design and highlight various interesting aspects of the detector readout electronics.

\subsubsection{Basics of OSR development with MUSIC}


   \begin{figure*}
   \begin{center}
   \includegraphics[width=16cm]{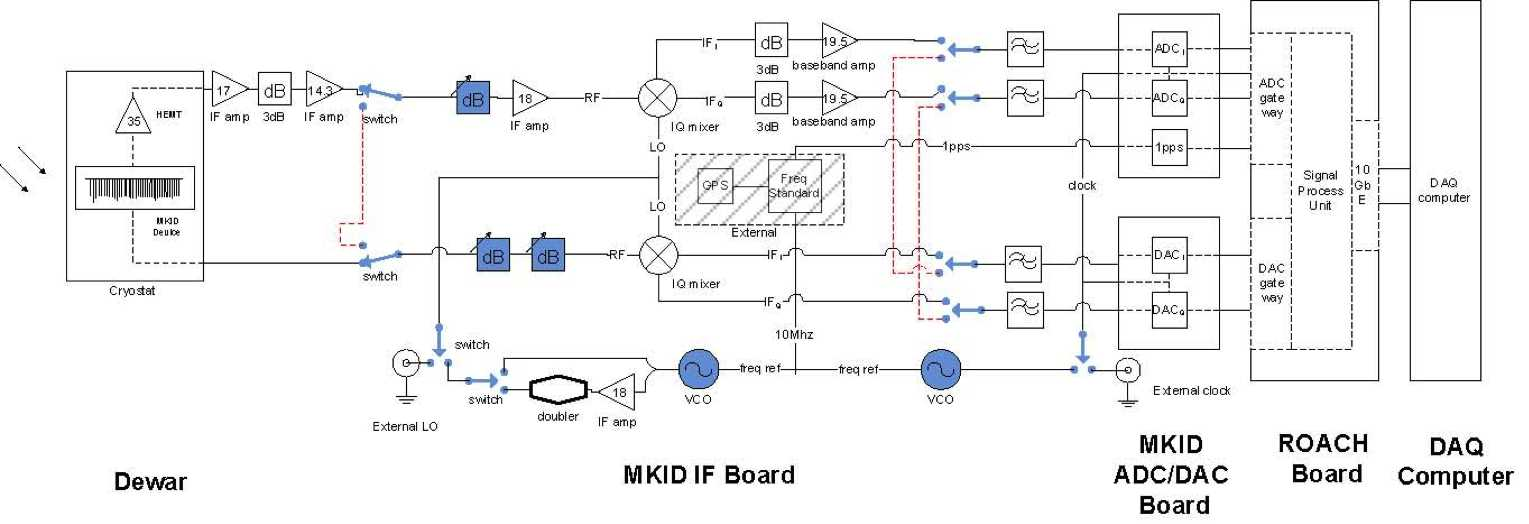}
   \end{center}
   \caption[Block diagram of the full readout system.]
   { \label{fig:Block diagram of the full readout system}
Block diagram of the full readout system.}
   \end{figure*}

    Fig. 6 presents an overview of the MUSIC readout system. The
      basic concept of resonator readout is to use IQ homodyne mixing,
      which is essentially a dual-phase lock-in detection
      technique. In general, the signal forms a closed loop with an FPGA to be the start and end points; along one path in the FPGA, the readout electronics send frequency tones to the device in the cryostat [the lower half of the diagram, which includes the digital-to-analog converter (DAC), IQ mixer, and digital attenuators], and along the other path [the upper half of the diagram, which includes the amplifier, attenuator, IQ mixer, and analog-to-digital converter (ADC)], the readout electronics receive the signal from the cryostat output and process the signal. Both signal paths utilize the DAQ computer, the signal processing board (an FPGA board known as Reconfigurable Open Architecture Computing Hardware [ROACH]\cite{parsons2006,Joshua,Parsons,Glenn,McMahon,mspec}, an ADC/DAC board, and an intermediate frequency board are shown in the diagram, which will be explained in detail in the following sections), and a cryostat.

   To read out the signal from the detector wafer, a comb of probe frequencies is generated for all resonators coupled to a single transmission line. This frequency comb is sent through the KID array, where the array modifies the amplitude and phase of the comb signals on the basis of the change in the surface impedance of the superconductor caused by the incident photons.

   After amplification by a cryogenic amplifier such as a high-electron-mobility transistor (HEMT) or silicon--germanium bipolar-junction transistor, the comb frequencies are transmitted to room-temperature electronics for digitization and analysis. There are no cryogenic components other than the cryogenic amplifier, attenuators, and KID wafer. For a resonator operating at a few gigahertz, a microwave probe signal can be generated by up-converting (mixing with a local oscillator) the baseband signal produced by sending a preprogrammed waveform stored on a memory block through a fast DAC card. Similarly, on the receiver side, the microwave signal is first down-converted by a mixer and then digitized by a fast ADC card. Recent advances in software-defined radio technology have provided additional options for fast signal processing. For example, the signal processing for a digitized signal can be performed digitally using algorithms on an FPGA.

   Prior to 2009, detector readout in our group was performed using off-the-shelf equipment, and only a few detectors could be read out. Between 2009 and 2010, we developed an open-source FPGA-based readout. We constructed prototypes of the DAC and ADC boards (Section II.C.2) and demonstrated the simultaneous readout of 126 detectors. From 2010 to 2012, we developed intermediate frequency (IF) boards (Section II.C.3), which helped us integrate the electronics and improve stability. Other advances included the second generation of the combined DAC--ADC board, an improved version of the firmware on an FPGA, expansion of the scale from 1 to 16 boards, and the production of a full set of DAQ software. The objective was to develop an OSR that could handle all of the tasks required for KID readout with a high level of automation.

   This section summarizes the work our group has performed with respect to readout electronics since 2009. In the summers of 2010 and 2012, we conducted two successful engineering runs at the CSO. The readout electronics have been intensively tested and are now in operation at the CSO for scientific observation.

\subsection{Readout design and development}

\subsubsection{Noise requirements and calculation}

For the majority of KID readout applications, the signal power at the detector is less than $-70$~dBm. Therefore, amplification is required in the signal-receiving chain before the room-temperature electronics. To ensure that the entire signal-receiving chain provides good signal-to-noise performance, a high-gain and low-noise component must be placed at the front of the signal-receiving chain. Cryogenic amplifiers (HEMT or SiGe) are the best components available for this purpose. In general, a cryogenic amplifier has a noise temperature of 2--5 K across the frequency band inhabited by our resonators. Therefore, the noise of the readout electronics is designed such that the white noise of the HEMT amplifier is the dominant contribution relative to the rest of the electronics noise.

   From the readout power of the carrier tone for each detector, we can calculate the signal-to-noise ratio (SNR) requirement at the HEMT. In our case, each resonator has a readout power between 10 and 30 pW at the device before the HEMT, meaning that the 144 resonators have a total power of $-58.4$~dBm. Inside the cryostat, the HEMT gain is $35$~dB, and a conservative value of the HEMT noise temperature, which places the most stringent requirements possible on the remainder of the electronics, is 2 K. The SNR at the HEMT output is approximately $56.68$~dB, as calculated in Section II.E.1. We carry this SNR throughout the following procedures.

   To develop the MUSIC readout system, we carefully designed the components to minimize the SNR degradation between the cryostat and the ADC to 1--2~dB, as discussed in Section II.H.5. The SNR requirement in front of the ADC is approximately $55.94$~dB for a 2-K HEMT noise temperature. We obtained suitable commercially available ADC chips that satisfied the SNR and bandwidth requirements with a margin of a few decibels.

\subsubsection{Bandwidth and center-frequency considerations}

The bandwidth of every OSR system is a tradeoff between the wafer design (how compactly the resonators can be packed with respect to the frequency) and the electronics limitations due to the SNR, the bandwidth, and the maximum number of detectors that the readout electronics can process. In our case, each resonator has a bandwidth of approximately 200--400 kHz\footnote{On the basis of the quality factor required for MUSIC, the coupling quality factors range from 60k to 110k. We found that Qc = 60k maximizes the mapping speed. Moreover, on the basis of a resonance frequency of 3 GHz (which is the KID resonance frequency that we knew how to design at that time), we determined that the full width at half maximum (FWHM) of the resonator was approximately 100 kHz \cite{jamesthesis}.}. We established a 2-MHz separation between the resonators, which is an intentionally large margin to account for shifts in the resonator position caused by fabrication errors. The relative positions of the resonators will not change during observation or cool-down cycles, but the fabrication variation can cause them to shift in position relative to one another.

   In the previous section, we determined that the ADC should have an SNR greater than $55.9$~dB. If we consider only quantization noise, then this SNR would theoretically require an ADC with at least 10 bits. However, the noise of fast ADCs is greater than the quantization noise expected on the basis of the number of bits; thus, in practice, we require a 12-bit ADC. The best 12-bit chip commercially available in 2009 has a 64-dB SNR and a sampling rate of up to 550 MSPS (chip model number TI ADS5486). Once we confirmed this choice of ADC chip, we expanded the separation between the resonators to 2.5 MHz.

   The center-frequency requirement for the readout is similar to the center-frequency requirement for the KIDs themselves and commonly ranges from 2 to 8 GHz. For MUSIC, the initial KID design had a resonator frequency of approximately 3--4 GHz, and the OSR system we developed is therefore also centered in that range. More recent KID designs can operate at frequencies as low as approximately 100 MHz. A lower frequency is attractive because it provides greater responsiveness in the frequency direction and thus lowers the two-level system (TLS) noise. Our group is also conducting extensive research on KIDs of this type.

\subsubsection{Readout of both in-phase and quadrature-phase components}

This section explains why we use two ADC and two DAC chips (which we often refer to as I and Q, respectively, because these chips are used with IQ mixers) and the relationship between the in-phase and quadrature-phase components.

   Regardless of the signal-processing method used, we are limited by the Nyquist sampling theorem. We use two chips to cover a wider bandwidth, thereby providing us with a useful bandwidth that is equal to the full sampling rate of 550 MHz rather than merely the Nyquist bandwidth of 275 MHz. If we only needed to read the amplitude or the phase information of the frequency tone, various DSP methods could be implemented, such as recovering the amplitude information from the data stream (ignoring the phase component) to enable the full 550-MHz bandwidth to be read using only one DAC and one ADC chip. However, in this case, phase information would no longer be available. In summary, two ADCs and two DACs are employed to obtain both the amplitude and phase information of the full sampling bandwidth.

   For detectors resonating at a few gigahertz, a mixer is required to convert the frequency from the baseband to the gigahertz band. An IQ mixer is a natural alternative to two single-sideband mixers for the simultaneous conversion of two signals. The use of an IQ mixer to utilize the full sampling bandwidth suggests that the baseband signal can be generated as DAC I and DAC Q, where I and Q have a phase difference of 90 degrees. In a complex data stream, I and Q serve as the real and imaginary parts of the complex data. If we perform a Fourier transform on a complex IQ data stream in the time domain, we obtain the full 550-MHz bandwidth in the frequency domain.

      The electronics emit (from the DACs) and receive (by the ADCs) the drive tones with both the amplitude and phase information\footnote{The amplitude and phase information can be captured from the in-phase and quadrature-phase components, respectively}. The submillimeter signal from the sky is converted into a change in the surface impedance of the superconducting inductor. We obtain the signal by monitoring the drive tones that couple with the resonators.

\subsection{Hardware}

\subsubsection{Overview of the OSR hardware, software, and firmware}

The OSR system can be divided into three parts: hardware (Section II.C), firmware (Section II.D), and software (Section II.F). In general, the hardware includes a customized ADC/DAC board, an IF board, an FPGA-based signal processing board, and an auxiliary system [e.g., a frequency standard or global positioning system (GPS)]. Firmware refers to the program that runs on the FPGA chip, whereas software includes all of the programming implemented to control and automate the readout. Each part of the OSR system will be discussed in detail in the following sections.

\subsubsection{ADC and DAC boards}

\paragraph{Chip selection.}

There are several qualified ADC chips on the market, and the SNR and
sampling-rate requirements for ADC chips have been discussed in previous
sections. In addition to these requirements, the spurious-free
dynamic range (SFDR) and intermodulation distortion (IMD) must also be
considered to prevent harmonics or spurs from impacting the SNR of the
resonator, particularly as the number of tones being read
increases. We selected chips with a random spur frequency power level
well below any level that could affect our resonators. For MUSIC, the
signal considered here is located in the range of 0.1--10 Hz near the
carrier tones. The resonators occupy a very small fraction of the
entire radio-frequency (RF) bandwidth; therefore, the probability of a spur
occurring within the resonator signal bandwidth is very low. However,
for some applications such as the KID dark-matter detector developed
at Caltech, the resonators must be monitored at higher frequencies (of
a few kilohertz) for pulse detection. In this case, the harmonics or
intermodulated frequencies are within a power level and frequency
range that could affect the detection results of the resonators; therefore, the IMD and harmonics must be considered when generating carrier tones, e.g., by designing the drive tones to avoid these harmonics and IMD frequencies.

   The choice of DACs is more flexible than the choice of ADCs; 16-bit and 1-GSPS DACs are readily and commercially available. We ultimately chose DAC5681 from Analog Devices, Inc. (Norwood, MA, US) and ADS5486 from Texas Instruments (Dallas, TX, US). The DAC can operate at up to 1 GSPS with a measured SNR of $75$~dBFS, and the ADC can operate at up to 550 MSPS with a measured SNR of $64$~dBFS. The 16-bit DAC and 12-bit ADC were both evaluated to confirm that they satisfied the SNR, SFDR, and IMD requirements. As a result of the rapid development of new semiconductor chips, faster ADCs and DACs have already appeared on the market, such as the 12-bit 3-GSPS ADC chip announced by Analog Devices, Inc. These semiconductor advances will be taken into consideration for future ADC/DAC development.

   The ROACH boards that we selected use Zdok connectors (as shown in Figs. 7 and 8) to connect to the ADC/DAC boards, allowing us to develop the ADC/DAC board independently and to use the ADC/DAC boards for different cameras and applications.

\paragraph{Clock and 1-PPS signals on the hardware boards.}

ADC and DAC boards require a clock frequency to operate. This clock frequency can be provided by either the FPGA or by an external clock source such as a standalone frequency synthesizer. An external clock provides the flexibility to choose any clock frequency and more stable performance (to satisfy high timing requirements) than that provided by an FPGA clock. We decided to use an external clock for both the ADC and DAC chips. We also chose to use the ADC's external clock for the FPGA to achieve better phase performance and synchronization with the ADC, DAC, and FPGA. We integrated the clock-frequency generation and local oscillator (LO) frequency generation on our IF board (which will be discussed in detail in Section II.C.3). The frequency stability originates from the 10-MHz reference, which is fed into the IF board and used to lock the clock and LO.

   To synchronize the system, we added a synchronization port to the ADC board. This port provides a link from an external one-pulse-per-second (PPS) signal generated by a GPS device to the FPGA fabric through the same Zdok connector used by the ADC and DAC boards.

\paragraph{Two generations of ADC/DAC boards.}

The low-frequency noise in the system is extremely important because we desire stability of the electronics system for 10 s or longer, which is not an important consideration for many room-temperature semiconductor manufacturers.

The first-generation board used independent voltage references for the ADCs and DACs, which added low-frequency noise arising from shifts in the two voltage references relative to one another. After testing the first-generation board, we also determined that it would not be possible to use the direct-current (DC) supply from the FPGA boards for the ADC/DAC board and that, instead, an external clean power supply would be necessary; moreover, the temperature of the ADC chips was found to increase quickly, which caused the low-frequency noise to worsen at high ADC chip temperatures.

We developed a second-generation ADC/DAC board with a common voltage reference (chip model ADR441), which substantially improved the low-frequency stability. The layout of this board is illustrated in Fig. 9. In addition, the new board was designed to have substantially improved thermal stability. Other advances included an external voltage supply with a regulator as well as modification of the positions of the SMA connectors to facilitate mating the ADC/DAC board to the IF board.

Figs. 7 and 8 show the layout of the MUSIC ADC/DAC board. The corresponding components are labeled in the layout. The figures show the following components of the board: two DAC chips, two ADC chips, a clock for the ADC, a clock for the DAC, the 1-PPS input, the Zdok connectors, and the common voltage reference.

\paragraph{DAC roll-off.}

The DAC output experiences frequency-dependent roll-off as a result of the following factors:

      1. The anti-aliasing filter, a Mini-Circuits low-pass filter (LPF) with a cut-off at 250 MHz and a DAC sampling rate of 500 MSPS. Its $S_{21}$ decreases from $0$~dB at 8 MHz to $-0.4$~dB at 200 MHz.

      2. The DAC chip itself. The output from even an ideal DAC suffers from the fact that the output is held constant for each clock cycle rather than smoothly changing. This causes the programmed DAC output to be modified by a sinc function in the frequency space. For a 500-MHz sampling rate, the sinc function causes $-0.004$~dB of attenuation at 8 MHz and $-1.516$~dB of attenuation at 200 MHz, with decreased attenuation for higher sampling rates.

      3. The transformer (which functions as a high-pass filter) on
      the DAC board, which causes the outputs to roll off below 5 MHz.

     The combination of the above factors means that the DAC output may experience a roll-off of approximately $2$~dB across the full bandwidth. To output well-behaved frequency tones to optimally drive the resonator on the camera wafer and produce the optimal sensitivity, the roll-off factors listed above were considered and compensated for by programming the carrier-tone power in the generation buffer accordingly.

  \begin{figure}[H]
  \includegraphics[width=8.0cm]{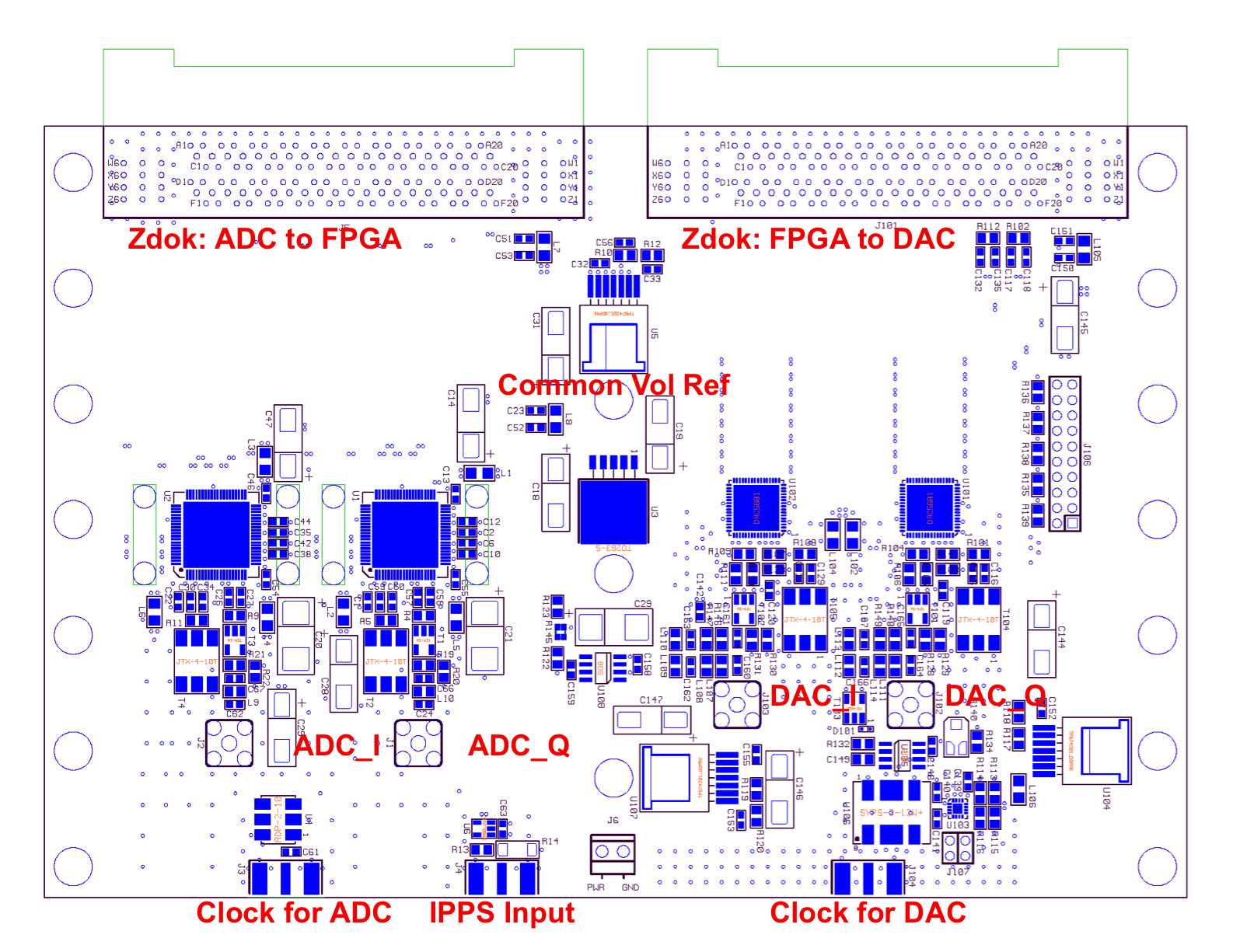}
  \caption[MUSIC ADC/DAC board layout design.]
  { \label{fig:MUSIC ADC/DAC board layout design}
MUSIC ADC/DAC board layout design.}
  \end{figure}

  \begin{figure}[H]
  \includegraphics[width=8.0cm]{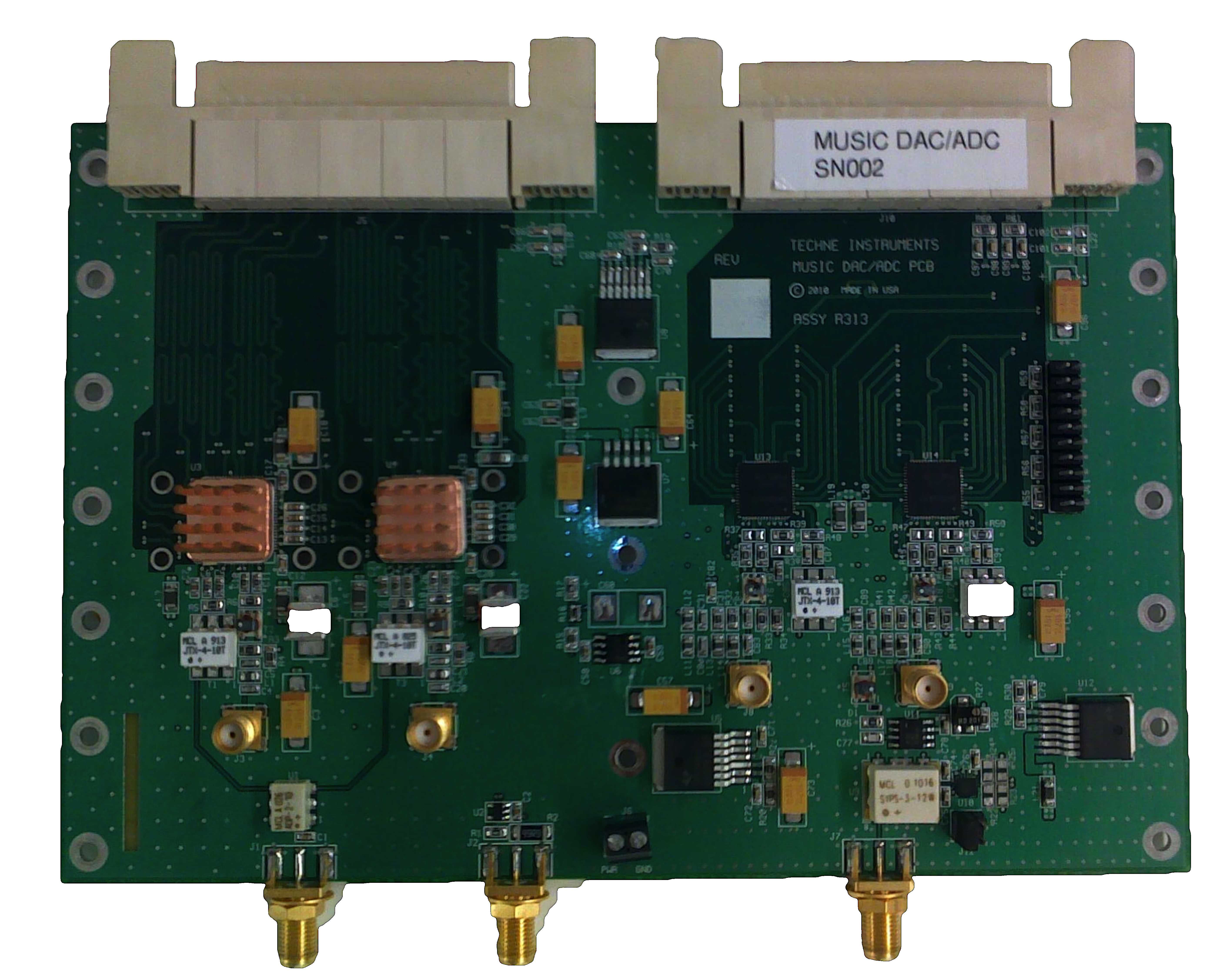}
  \caption[Photograph of the MUSIC ADC/DAC board.]
  { \label{fig:Picture of MUSIC ADC/DAC board}
Photograph of the MUSIC ADC/DAC board.}
  \end{figure}

  \begin{figure}[H]
  \includegraphics[width=7.5cm]{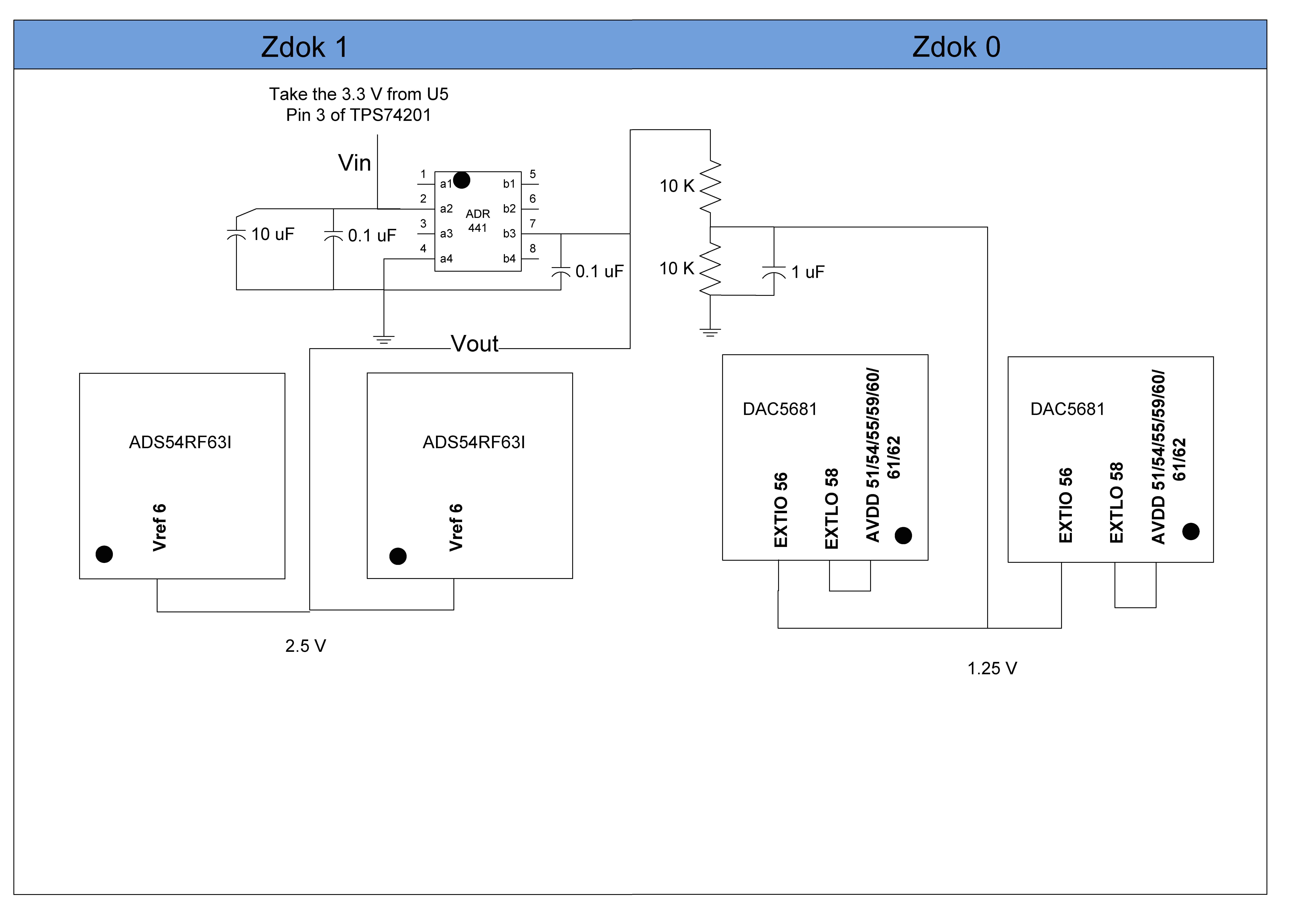}
  \caption[Common voltage reference for the ADC/DAC board.]
  { \label{fig:Common voltage reference for ADC/DAC board}
Layout of the common voltage reference for the ADC/DAC board.}
  \end{figure}

\subsubsection{IF board}

Because the DAC and ADC work with baseband signals (DC to the Nyquist
sampling rate) and because our resonator operates at microwave frequencies (a few gigahertz), the baseband signals must be up-converted to or down-converted from the resonator frequency band. The basic hardware components of the OSR system are described in Fig. 6 and Section II.A.

After the DAC board emits the frequency carrier signal, the signal passes through an LPF, an IQ up-converting mixer, and a digital attenuator before entering the dewar. On the receiver side, the carrier signal from the device is amplified by the HEMT and by room-temperature low-noise amplifiers (LNAs) before passing through a digital attenuator and being down-converted by another IQ mixer to the baseband. The down-converted carrier signal is then amplified, Nyquist filtered, and finally digitized by the DAC.

Many integrated circuit (IC) components are used in the intermediate stage to convert the frequency, control the power level, and perform a loopback test. The purpose of an IF board is to integrate all of the hardware components between the ADC/DAC board and the dewar.

In 2011, we designed and fabricated the IF board. Figs. 10 and 11 present the IF board fabrication layout, which contains the following:

 \begin{itemize}

\item two voltage-controlled oscillators (VCOs), which provide the FPGA clock signal (ranging from 137.5 to 4400 MHz) and the IQ mixer LO signal (ranging from 2.2 to 4.4 GHz or, with the doubler, from 4.4 to 5 GHz);

\item two IQ mixers, which convert between the baseband and resonator frequency ranges;

\item two digital attenuators (each with attenuation from $0$ to $31.5$~dB in 0.5-dB steps) to set power levels for transmission into the dewar; one digital attenuator and five amplifiers that serve the function of setting power levels for reception by the IQ mixer, and reception by the ADC. 

\item   nine digital switches that allow the signal to be looped back
  at various points in the signal chain (e.g., baseband loopback,
  which bypasses the entire up-conversion/down-conversion and dewar
  chain, and RF loopback, which bypasses the dewar) and provide the option of choosing an external clock, an external LO, or doubling the LO frequency;

\item one frequency doubler to double the LO frequency for higher resonant-frequency applications; and
\item  four anti-aliasing LPFs.
 \end{itemize}

  \begin{figure}[H]
  \includegraphics[width=8.5cm]{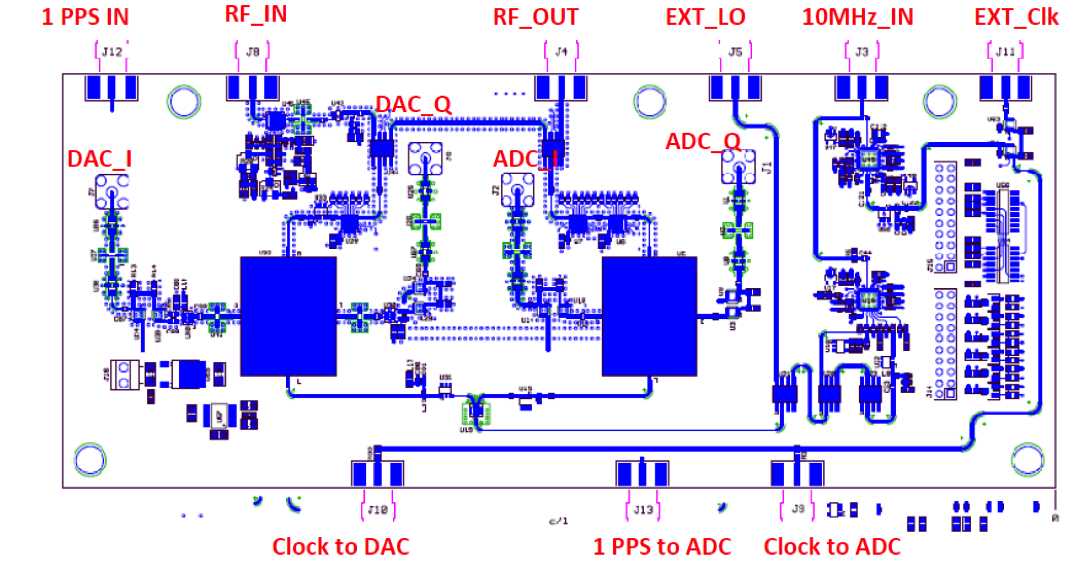}
  \caption[MUSIC IF board layout.]
  { \label{fig:MUSIC IF board layout}
MUSIC IF board layout.}
  \end{figure}

  \begin{figure}[H]
  \includegraphics[width=8.5cm]{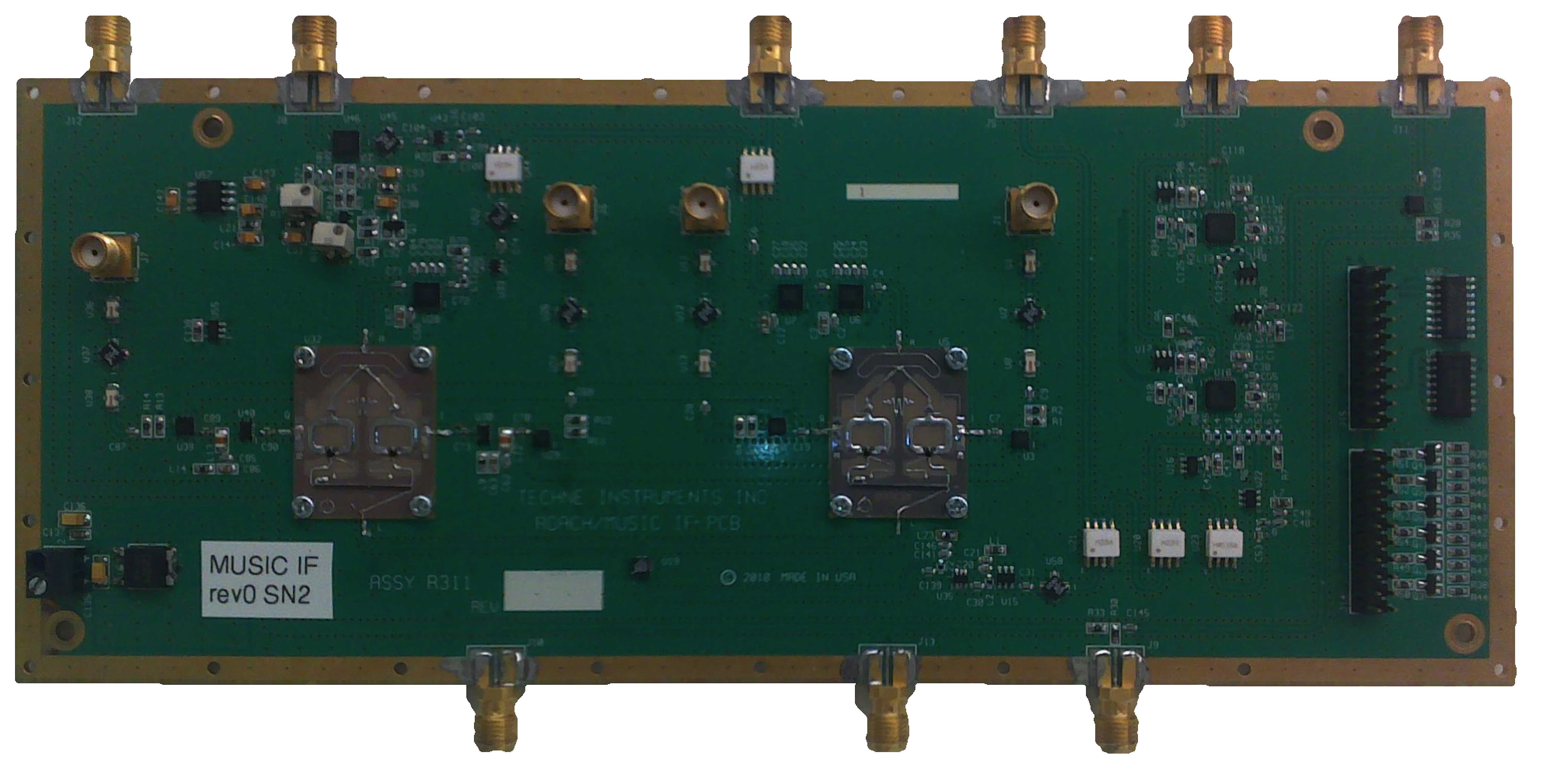}
  \caption[Photograph of the MUSIC IF board.]
  { \label{fig:Picture of MUSIC IF board}
Photograph of the MUSIC IF board.}
  \end{figure}

   All of these functions of the IF board can be digitally controlled
   or programmed by the ROACH board. The control signal is sent from
   the FPGA and connected to the GPIO connector on the IF board via a
   20-pin ribbon cable. The control firmware is designed to be
   independent and can be added to any existing firmware; thus, the IF
   board can be fully controlled or reprogrammed while the
   channelizing firmware is running. Each component on the IF board
   was carefully selected and configured to ensure that the noise level that reaches the ADC is dominated by the HEMT noise, and all other components in the system, such as the amplifiers and ADC, add a negligible amount of noise. By virtue of the IF board and a carefully designed DAC buffer, the probe signal that reaches the MKID device is optimized for each individual resonator across the entire readout bandwidth in both frequency and amplitude.

   A noise study of the IF board will be presented in Section
   II.H.5. The details of the IF board test are described in Appendices A and B.

\subsubsection{FPGA board}

The FPGA processing board that we used was developed by the Collaboration for Astronomy Signal Processing and Electronics Research (CASPER) and is a standalone FPGA board known as ROACH (three connected circuit boards are shown in Fig. 12; the ROACH board is on the left, the ADC/DAC board is in the middle, and the IF board is on the right. All three boards are mounted together on an Al plate to prevent relative vibration and flexing of the connections and to provide a good heat sink for the system).

   The central component of the ROACH board is the Xilinx Vertex 5 FGPA. An independent PowerPC (PPC) running the Linux operating system is used to control the FPGA. In addition to the memory on the FPGA, one double data rate (DDR) dynamic random-access memory (DRAM) module and two quad data rate (QDR) static RAM modules are used to provide extra memory capacity to store the lookup table (LUT; this is the buffer played back by the DAC), provide the fast Fourier transform (FFT) coefficients, and buffer intermediate data during FPGA signal processing; two Zdok connectors allow the DAC, the ADC, or other interfaces to attach to the FPGA; four CX4 input and output connectors provide a data rate of up to 40 Gbps to transfer data or link multiple ROACHes together; and a 1-Gbps Ethernet port can be used by the DAQ computer for data communication and control of the PPC on the board \cite{roach}.

  \begin{figure}[H]
  \includegraphics[width=8.5cm]{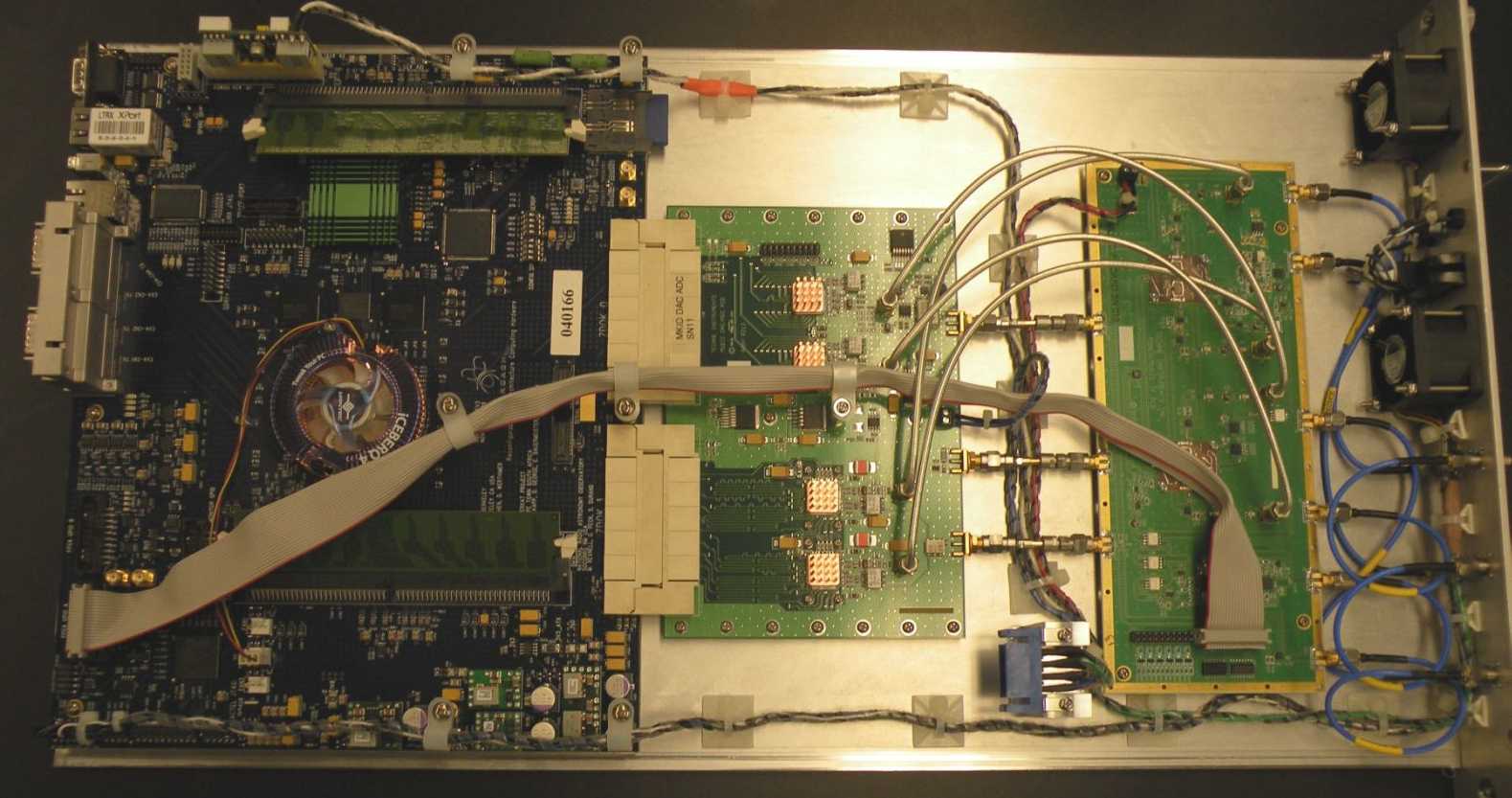}
  \caption[Photograph of the readout hardware mounted on the Al plate.]
  { \label{fig:Picture of the readout hardware mounted on the Al plate}
Photograph of the readout hardware mounted on the Al plate.}
  \end{figure}

\subsubsection{Auxiliary system}

As previously mentioned, the readout electronics require two auxiliary signal inputs: a 1-PPS signal and 10-MHz frequency reference. The IF board directly uses the 10-MHz standard to generate the LO and FPGA/ADC/DAC clocks. The 1-PPS signal is routed through the IF board and ADC/DAC board to the ROACH board to be used for time synchronization. A GPS receiver (used by the entire CSO; model number Garmin LVC 18x) generates a 1-PPS signal. This 1-PPS signal is received by a 10-MHz rubidium frequency standard (Stanford Research Systems [SRS] FS725). The 10-MHz standard is locked to the 1-PPS signal. It averages over a long timescale and outputs a stabilized 1-PPS signal (as well as the 10-MHz reference). We use distribution amplifiers (SRS FS735/1/1 for the 10-MHz signal and SRS FS735/3/3 for the 1-PPS signal) to generate 16 copies of the 1-PPS and 10-MHz signals, which are routed to the individual OSR units. We use the GPS-sourced 1-PPS signal to ensure precise synchronization with the telescope pointing timestream and precise knowledge of the absolute time. Details of the synchronization of the time samples to an absolute time reference will be discussed in Section II.D.5.

\subsection{FPGA firmware}

The previous section described the hardware in the system. This section will discuss the development of the firmware for the OSR (i.e., the program running on the FPGA chip).

      A flow chart that provides an overview of the firmware behavior is presented in Fig. 13. The lower diagram shows the DAC LUT and DAC gateway, which is the direction in which the FPGA plays back the LUT buffer and sends the signal to the DAC board. This pathway includes a serialization step to provide the DAC with four DAC clock cycles of data during each FPGA clock cycle. The upper diagram shows the signal receiving and processing chain on the FPGA. On the receiving side, the ADC outputs the data in parallelized (deserialized) format, i.e., four ADC clock cycles for each FPGA clock cycle. The corresponding data streams from the I and Q ADCs are added to obtain a complex number, and the four complex streams are then processed by four parallel FFT cores (housed in two Biplex FFT blocks). Of the $2^{16}$ bins output by the four parallel FFTs, we are interested only in the 192 that have carrier tones; therefore, the data stream passes through a frequency-bin selector, and then, the 192 data streams pass through a 192-channel FIR filter to decimate the data rate by a factor of 75, which makes the final data rate exactly 100 Hz. Before the data are sent out of the FPGA, we package the data stream into a transmission control protocol (TCP) packet with a timestamp.

  \begin{figure*}
  \begin{center}
  \includegraphics[width=14cm]{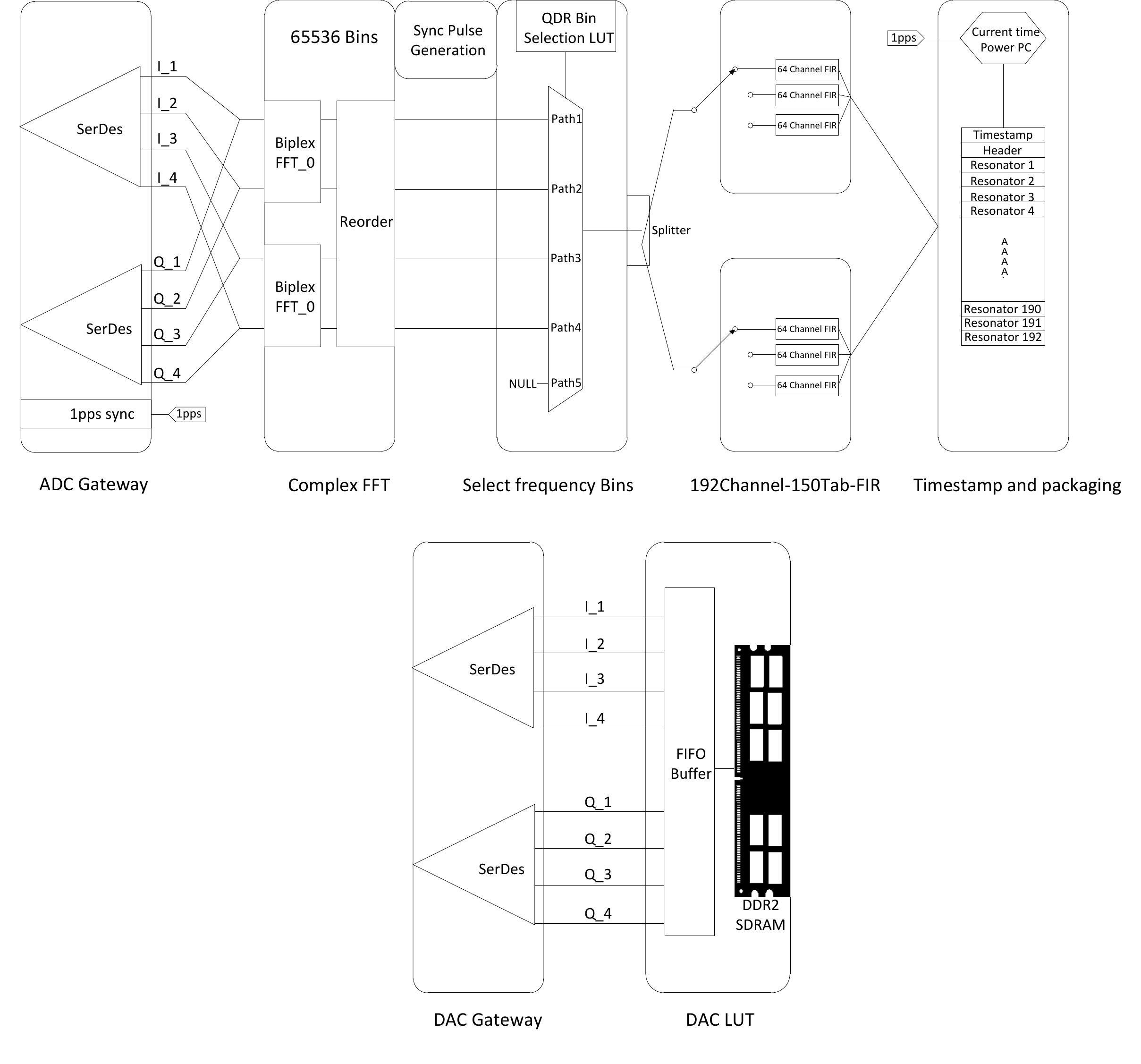}
  \end{center}
  \caption[Flow chart of the FPGA firmware.]
  { \label{fig:Flow chart of FPGA firmware}
Flow chart of the FPGA firmware: (top) from the ADC input to the FPGA output and (bottom)
lookup table and DAC.}
  \end{figure*}

\subsubsection{Signal-Tone LUT generation}

The DAC plays back a stored LUT to send out a waveform that consists of approximately 200 sine waves with predefined frequencies and phases. This section describes how this LUT is generated.

\paragraph{Frequency-Tone resolution and LUT storage.}

The length of the LUT is defined by how precisely we need to drive the resonator tones. The criterion is that the drive tone should match the resonance frequency to a precision of 10\% of the resonator FWHM bandwidth. The on-sky resonator quality factors of 60k to 110k combined with resonance frequencies of 3--4 GHz imply a resonator FWHM of approximately 100 kHz; therefore, a precision of 10 kHz is required.

To obtain consistent phase information for each readout tone, the DAC frequency-bin width must be equal to or an integer multiple of the FFT channelizer resolution. In our case, a bandwidth of approximately 500 MHz with a $2^{16}$-bin FFT yields a resolution of approximately 7.5 kHz. Because we will later decimate the data to the CSO-telescope pointing update rate of 100 Hz, the output data rate and thus the frequency resolution of the FFT should be a multiple of 100 Hz; hence, we choose a 7500-Hz FFT resolution rather than $500/2^{16}$ MHz. A total of $2^{16}$ FFT bins yield a 491.52-MHz bandwidth.

The most convenient place to store the LUT is in the FPGA fabric. However, there is limited memory space on the FPGA, and all of that space is required for the FFT. Instead, we successfully stored the LUT in the DRAM while satisfying the LUT synchronization performance requirements by carefully tuning the latency between the DRAM input and the output addresses.

\paragraph{Roll-Off pattern compensation.}

The full transfer function of the readout chain, including the effects of the DAC intrinsic sinc function, LPFs, IQ mixers, transformers, impedance mismatches, standing waves, cable delay, etc., must be considered when setting the optimal power level for each resonator. We used the network analyzer mode (described in Section II.D.6) to record the roll-off pattern of each frequency bin for the full sampling bandwidth in the current system setup. We compensated for the roll-off by adjusting the DAC LUT and digital attenuator to optimize both the power level and the frequency across the entire readout bandwidth for each resonator in the MKID array.

\paragraph{Real-Time reprogramming.}

During KID camera observation, the resonance frequency changes when the optical power on the resonator changes (e.g., when the telescope points toward a different part of the sky or when the atmospheric opacity changes). To compensate for these changes, we designed the firmware to allow for rapid updating of the DAC LUT using a new buffer with new drive tones while the channelization firmware on the FPGA continues to run (it takes less than one second for a new set of LUT buffers to upload).

\paragraph{Anticlipping algorithm.}

The MKID LUT waveform is the summation of approximately 200 different carrier frequencies of similar but different powers. In an actual DAC, this summation poses a clipping problem, namely, the sum of the carrier values at a certain waveform position might exceed the maximum range of the DAC, even when the average power is within the DAC limits and when the carrier phases are randomly defined. Simply scaling the waveform amplitude to fit within the DAC limits will result in a reduction in the carrier and average power by the square of the scaling factor. This problem is often described in terms of the peak-to-average-power ratio (PAR). The SNR of the DAC is not effectively utilized when the PAR is significantly greater than unity.

Under the assumption of a sinusoidal wave output from a DAC with $N$ bits, the theoretical maximum SNR can be calculated using the following quantization noise formula: SNR = ($1.761 + 6.02 \times N$) dB. From experience, if we simply use random phases for the 200 carrier tones with an LUT length of $2^{16}$, we will obtain an SNR that is 1--2~dB below the theoretical maximum SNR defined by the effective number of bits. However, if we use a random phase, the summation of 200 carrier tones in the time domain may overflow the maximum range that can be defined using the number of bits available. Using MATLAB's random function to generate the phases, we must conduct many trials to ensure that no clipping occurs (occasionally as many as 100 trials). The indeterminacy of such an algorithm is undesirable for a scientific instrument, and this algorithm does not scale well with the number of carriers.

Another possible solution is to simply ``clip'' the waveform at the DAC limits, regardless of the consequences. Clipping the waveform does not strongly affect the average power, but it scatters power from the carrier frequencies into other frequency bins. This effect can be understood by equating the clipping of the waveform to the addition of a scaled delta function to the time sample at which the clip occurs and from the fact that such a pulse has a white frequency spectrum. The PAR in this scheme is better than that in the scaling scheme, but the power that remains in any given carrier will differ from what was originally designed.

This discussion suggests an alternative approach: at time samples at which the waveform would clip, subtract off a pulse that is designed to reduce the waveform amplitude below the DAC range but whose Fourier spectrum has had the carrier-frequency bins removed. In this manner, a signal (or noise) is introduced into the noncarrier bins in a manner similar to the introduction of a higher white-noise level, but this noise does not affect the SNR for the carrier bins. The applicability of this algorithm ultimately requires a high ratio of noncarrier bins to carrier bins. This clipping avoidance program can generate waveforms with a PAR approaching unity, making essentially full use of the dynamic range of the DAC.

\subsubsection{ADC digitization}

      A clock rate of approximately 500 MSPS is too fast, even for a state-of-the-art FGPA. To process the data from the ADC and DAC in real time, we first deserialize the data (Table 1) on the FPGA to reduce the FPGA clock rate by a factor of four (e.g., the clock rate on the FPGA becomes 491.52/4 MHz) at the expense of requiring the FPGA to simultaneously process four parallel data streams\cite{parsonsfft}.

 Furthermore, FFT algorithms have been designed to handle this parallelization of the data streams, as will be described in the next section. If we consider the $2^{16}$ nominal output bins of a nonparallelized FFT and number them sequentially as 1, 2, 3, etc., the parallelized FFT algorithm will then output four parallel streams with the frequency bins numbered 1, $2^{14}+1$, $2 \times 2^{14}+1$, $3 \times 2^{14}+1$ output on one clock cycle; bins 2, $2^{14}+2$, $2 \times 2^{14}+2$, $3 \times 2^{14}+2$ output on the next clock cycle; and so on.

\begin{table*}
\caption{Deserialization of the data stream into four parallel paths for the FFT in the FPGA.}
\label{lab:2}
\begin{tabular}{cccccccc}
\hline
FPGA Clock Cycle &  Cycle 1 & Cycle 2 &  $\ldots$ & Cycle 16383&  Cycle 16384 \\
FFT O/P 1 & Bin1 &  Bin2 &  $\ldots$ &  Bin $2^{14}-1$ &  Bin $2^{14}$\\
FFT O/P 2 & Bin $2^{14}+1$ & Bin $2^{14}+2$ & $\ldots$ & Bin $2 \times 2^{14}-1$ &  Bin $2 \times 2^{14}$\\
FFT O/P 3 & Bin $2 \times 2^{14}+1$ & Bin $2 \times 2^{14}+2$ & $\ldots$ & Bin $3 \times 2^{14}-1$ &  Bin $3 \times 2^{14}$\\
FFT O/P 4 & Bin $3 \times 2^{14}+1$ & Bin $3 \times 2^{14}+2$ & $\ldots$ &  Bin $4 \times 2^{14}-1$ & Bin $4 \times 2^{14}$\\
\hline
\end{tabular}
\end{table*}

As the requirements for the bandwidth and ADC sampling rate increase, the data streams and FFTs can be further parallelized if necessary at the expense of requiring larger or multiple FPGAs.

\subsubsection{Design of the frequency demultiplexer}

There are several methods for implementing a demultiplexer in an FPGA: a digital down-converter (DDC), a single-stage FFT, a polyphase filter bank (PFB), an FFT zoom (two cascaded FFTs), a combination of a PFB and an FFT, a combination of an FFT and a DDC, and a combination of coaddition and an FFT. These options were all considered and compared for the design of the demultiplexer. For example, a DDC would not be an appropriate choice for an increasing number of resonators because it requires a long storage waveform (compared with the length of a single-stage FFT) to play back for the readout of each carrier tone. An FFT zoom program will suffer significant power leakage in the first FFT stage, particularly when the frequency does not fall at the centers of the bins of the first FFT.

In 2010, we implemented a combination of a $2^{11}$-sample polyphase filter bank and a $2^{6}$-sample FFT for our first DemoCam run. We obtained better performance using the final design that was implemented for the 2012 commissioning runs, which used a single-stage $2^{16}$-sample FFT.

\paragraph{Design of the first-version firmware: PFB and FFT.}

The PFB was developed to account for the fact that the discrete Fourier transform (DFT) is an imperfect approximation of a set of BPFs. Ideally, each sample in a DFT would represent the response of a perfectly rectangular BPF centered on an integral multiple of $F_n/N$ (the Nyquist frequency divided by the size of the DFT) with a width of exactly $F_n/N$. In actuality, each sample in a DFT has a sinc frequency response. The PFB squares off this filter shape by first convolving the input with a sinc function over a long time window and then applying a DFT. The length of
this time window determines the steepness of the edges on the filter; typically, it may be four to eight times (four to eight taps of the filter) the length of the DFT.

In the combined PFB and FFT design used in the 2010 DemoCam run, we used a four-tap $2^{11}$-point Hamming-window complex polyphase filter bank followed by a transpose function implemented on the QDR and then a $2^{6}$-point complex FFT. This design is similar to an FFT zoom, but the use of the PFB substantially reduces the bin-to-bin power leakage present in the first FFT of an FFT zoom. A $2^{11}$-point PFB together with a $2^{16}$-point FFT yields $2^{17}$ frequency bins.

Compared with a direct $2^{17}$-point FFT, the combined PFB and FFT solution required considerably less space on the FPGA; a $2^{17}$-point FFT would not have fit. This design would have allowed us to expand the size of the channelizer up to 16 million points on a single ROACH (or even larger by combining ROACHes together through a 10-Gbit connector). The cascaded design shifts the FPGA fabric requirement for the FFT into the external QDR memory by dividing the large FFT into two parts.

The combined PFB and FFT design suffers intrinsic power leakage between bins in the first stage unless an infinitely long filter bank (impossible in practice) is implemented in the first stage. In a real FPGA implementation, the first-stage PFB is very expensive in terms of resources because it must sum and buffer the full raw timestream $X$ times, where $X$ is the number of taps in the PFB. The value of $X$ used is a tradeoff between the amount of FPGA fabric used and the quality of the response function.

   In the design that we used in 2010, we implemented a four-tap
   polyphase filter bank in the first stage (buffering the raw data
   stream four times). Because the PFB and FFT are both well-defined
   linear calculations, the bin-to-bin leakage is known and can be
   corrected for. However, doing so increases the amount of FPGA
   fabric required because additional frequency bins must now be
   carried along for every carrier bin. This method has been
   implemented and validated for a situation in which the number of
   carriers is not large. However, as the number of carriers
   increases, even the summation of only four additional bins for each carrier
   bin incurs considerable computational effort. In principle, this
   problem can be solved through matrix inversion, but this is also
   computationally challenging. One method to simplify the summation
   is to use a large first-stage PFB and a small second-stage
   FFT. However, this approach causes the two-stage design to become
   more similar to a large single-stage FFT, which conflicts with the original intent of conserving computing resources by using the two-stage design. One effect that must be accounted for in the PFB--FFT design is that the PFB induces a phase rotation of the Fourier coefficients. This rotation is deterministic and can be corrected for.

For radio astronomy applications other than KID readout that do not involve a large number of predefined carriers tones, the aforementioned power leakage problems are less important. For KID readout, however, they are an issue of considerable concern, and the difficulties will only increase as the number of tones per unit bandwidth increase. Ultimately, we chose to focus on implementing a large single-stage FFT.

\paragraph{Design of a large complex FFT on a Xilinx FGPA.}

We evaluated FFT designs from both the CASPER group and Xilinx. The CASPER FFT uses our four parallel timestreams exiting the ADC [termed ``Decimation In Time'' (DIT)]. The Xilinx FFT is more logic-cell efficient than the CASPER FFT for a large FFT size (greater than $2^{15}$) and a large output bit width (greater than 18 bits). We considered the advantages of both designs and ultimately developed an improved FFT design by recognizing the following deficiencies of the CASPER FFT block:

\begin{enumerate}
\item For digital processing, the number of bits required to store the number inside the FPGA for each stage must be determined to ensure that no information is lost during digital signal processing. We can address this issue from an SNR perspective. We know that the maximum possible SNR that can be represented by $N$ bits is ($1.761+6.02 \times N$)~dB. For example, if we use a $2^{16}$-sample FFT and a $64$-dB SNR at the ADC, the SNR at the FFT output will then be $112$~dB. Thus, the output must be at least 18.39 bits. Here, we use 20 bits for the FFT output and 19 bits for the calculation coefficients within the FFT process. A 20- or 19-bit output and coefficient calculation does not mean that no number can be greater than 20 bits during the FFT calculation. The intermediate numbers inside the FFT calculation are floating points and can thus be scaled on the basis of the calculation requirements.

For simplicity, the CASPER FFT assumes that the output bit width is
identical to the input bit width. However, the ADC does not have a
significant number of bits (e.g., our ADC has 12 bits), and this
method wastes FPGA resources if we use 19 bits for both the input and output of the FFT.

\item The CASPER FFT is constructed around biplex blocks that process two parallel input timestreams. For four parallel streams, it uses two biplex blocks. The coefficients used by all of these biplex blocks are identical, but each block possesses its own copy of the coefficients by default. By sharing a single copy of the coefficients, substantial FPGA fabric can be saved.
\item Some of the coefficients for the FFT calculation are stored in read-only memory (ROM) as an LUT for the CASPER FFT. However, those numbers can be generated on the fly using FPGA logic, thereby conserving a significant amount of resources.
\end{enumerate}

   To overcome the aforementioned deficiencies, we implemented an improved FFT block to solve the problems and to enable a single $2^{16}$-bin single-stage FFT design on a Vertex 5 FPGA. Table 2 summarizes the FPGA logic-cell utilization of the $2^{16}$-bin single-stage FFT alone (without other firmware functions running on the FPGA)\footnote{The default number of flip-flops on the FPGA is limited, but it can be replaced with slices when additional flip-flops are required.}.

\begin{table}[H]
\caption{FPGA logic-cell utilization summary.}
\label{tab:fonts}
\begin{tabular}{|c|c|c|c|c|}
\hline
\rule[-1ex]{0pt}{3.5ex}  Logic Utilization & Used & Available & Use Rate \\
\hline
\rule[-1ex]{0pt}{3.5ex}  Number of slice registers & 7348 & 58880 & $12\%$ \\
\hline
\rule[-1ex]{0pt}{3.5ex}  Number of slice LUTs & 5928 & 58880 & $10\%$ \\
\hline
\rule[-1ex]{0pt}{3.5ex}  Number of slices as memory & 2881 & 24320 & $11\%$   \\
\hline
\rule[-1ex]{0pt}{3.5ex}  Number of route-thrus & 329 & 117760 & $1\%$ \\
\hline
\rule[-1ex]{0pt}{3.5ex}  Number of flip-flop pairs used & 7575 & 7575 & $100\%$ \\
\hline
\rule[-1ex]{0pt}{3.5ex}  Number of bonded IO buffers & 448 & 640 & $70\%$ \\
\hline
\rule[-1ex]{0pt}{3.5ex}  Number of RAM/FIFO blocks & 20 & 244 & $8\%$ \\
\hline
\rule[-1ex]{0pt}{3.5ex}  Number of DSP48Es & 62 & 640 & $9\%$  \\
\hline
\end{tabular}
\end{table}

\subsubsection{Bin selection and decimation}

\paragraph{Bin-Selection design.}

Because we are only interested in the information contained in the approximately 200 bins that carry the resonator tones, we wish to extract these 200 bins from the full FFT. When the DAC LUT is generated, we also generate a position table for the FFT bins that carry resonator tones, which is automatically updated as the DAC LUT is updated while the FPGA channelization program is running.

   As previously discussed in the context of the ADC output streams (Section II.D.2), the signal timestream is deserialized into four parallel streams, and the FFT on the FPGA also has four parallel outputs (as summarized in Table 1).

We select bins by masking all output cycles with an LUT of $2^{16}/4$ in size. For each FPGA clock cycle, we determine whether there were any resonators within this clock cycle and, if so, which of the four parallel outputs are contained the resonator bin.

   Because there are four parallel simultaneous outputs for each clock cycle, we cannot select bins from two or more of the four parallel FFT outputs at once without buffering the four parallel output data streams. To bypass this difficulty, we ensure that no more than one FFT bin is selected per clock cycle. This scheme requires that two resonator tones not be separated by an exact multiple of $2^{14}$ bins (e.g., bin 1 and bin $2^{14}+1$ cannot be selected at the same time).

\paragraph{Decimation for astronomical signals.}

After selection of the resonator bins, approximately 200 data streams carrying resonator information are further processed and stored. The data rate after the FFT is 7500 Hz (the sampling rate divided by the FFT length). To reduce the data rate, rather than using simple coaddition or averaging, each selected resonator data stream is processed with a 150-tap Hamming-window FIR filter. The output timestream is then decimated by a factor of 75 to 100 Hz to match the update rate of the CSO-telescope pointing timestream. The use of an FIR filter prior to decimation provides better frequency--space characteristics than does coaddition or averaging. The signal flow for the decimation process is presented in Table 3. The selected carrier tones are assigned to FIR filters 1, 2 and 3 in a cyclical manner.

\begin{table*}
\centering 
\caption{Signal flow of the 192 resonators feeding into three FIR filters.}
\label{lab:2}
\begin{tabular}{cccccc}
\hline
- & Channel 1 & Channel 2& $\ldots$ & Channel 63 &  Channel 64\\
I/P to FIR 1 &  $Res\_1$ &  $Res\_4$ &   $\ldots$ & $Res\_187$ &  $Res\_190$\\
I/P to FIR 2& $Res\_2$ &  $Res\_5$ &   $\ldots$ & $Res\_188$ &  $Res\_191$\\
I/P to FIR 3& $Res\_3$ &  $Res\_6$ &   $\ldots$ & $Res\_189$ &  $Res\_192$\\
\hline
\end{tabular}
\end{table*}

As previously mentioned, there are 144 resonators within one $6\times6$ antenna array. Hence, there are 144 on-resonator frequency bins that contain resonator information. The extra readout capacity enables us to generate 50--60 tones in off-resonance frequency bins and fully read out those off-resonance carrier bins. We use these data to measure the common-mode electronics 1/f noise (HEMT and room-temperature electronics; this noise is discussed in detail in Section II.H.3).

We eventually chose 48 off-resonator carriers because each FIR filter block accepts 64 streams of data; thus, the three FIR blocks that are required to handle 144 resonators can process an additional 48 off-resonance streams without requiring additional FPGA resources.

For each channel, the 150 coefficients are stored using 22 bits because the SNR that is required at the FIR output for each channel is $130.9$~dB\footnote {As discussed in Section II.D.3.b, the SNR requirement at the FFT output is $112$~dB; after the FIR, with a decimation rate of 75, the SNR requirement becomes $112 + 10 \times \log_{10}(75)$~dB.}. At least 21.46 bits are needed for the coefficients to provide this output SNR.

   Here, we used the Xilinx FIR block. In the future, when a larger number of resonators must be read out and when the required number of FIR blocks cannot fit on the FPGA, the external memory (e.g., QDR or DRAM) can be used to store the FIR coefficients and buffer the data streams.

\subsubsection{Synchronization}

\paragraph{Timestamps and data packaging.}

The 1-PPS TTL signals are received on the IF board and routed through the ADC/DAC board and ROACH board to the FPGA, where they provide a logic signal with a rising edge on the exact-second boundary with respect to the absolute time of day. Both the DAC and channelizer begin at the same second boundary to ensure that we obtain a consistent phase for all carrier bins. For synchronization with the absolute time of day, a C program running on the PPC transfers the current Linux time from the PPC [which is locked to a network time protocol (NTP) server] to the FPGA. A counter in the FPGA that is locked to the 1-PPS signal begins counting in integer seconds from that time. Another counter on the FPGA serves to count the number of FPGA clock cycles since the last second boundary. Thus, the two counters combined provide timestamp information accurate to 1/(FPGA clock rate) $\approx$ 10 ns.

Each output data packet contains a timestamp (both seconds and fractional seconds), a header (a predefined number to separate the timestamp and data), and 192 complex resonator signal values (as summarized in Table 4). The data are transmitted via 1-Gbit Ethernet at 100 Hz to a DAQ computer using a TCP data server that we constructed for this purpose. The first-version firmware used 10-Gbit user datagram protocol (UDP) Ethernet, but this Ethernet card is expensive, and 16-port 10-Gbit Ethernet switches are not readily available. In addition, the UDP packet, although simple to set up, does not have handshake functions to protect against data loss. Thus, we switched to 1-Gbit Ethernet with TCP packets. After optimization of the connection parameters, we observed no lost packets when using TCP at the required data rate.

\begin{table*}
\caption{Output packet data format for MKID readout.}
\label{tab:fonts}
\begin{center}
\begin{tabular}{|l|l|l|l|l|l|}
\hline
\rule[-1ex]{0pt}{3.5ex} Position & 1 & 2 & 3 & ... & 194\\
\hline
\rule[-1ex]{0pt}{3.5ex}  top 32 bits&Unix seconds&header&res1 real part&...&res192 real part\\
\hline
\rule[-1ex]{0pt}{3.5ex}  bottom 32 bits&fractional second&header&res1 imaginary part&...&res192 imaginary part \\
\hline
\end{tabular}
\end{center}
\end{table*}

\subsubsection{Network analyzer mode and IQ sweep mode}

In addition to the mode described above, which is used for observations, we designed several other modes of firmware operation. The following are two commonly used firmware operation modes:

\begin{itemize}

 \item  IQ sweep mode. This mode is a variant of the normal DAQ mode. Normally, the LO is held fixed while data are acquired during an observation. Instead, we acquire short blocks of data (on the order of 0.5--1 s) in this mode to characterize the system's complex transmission as a function of the frequency near each resonance, and between blocks, we step the LO away from its nominal value over a range that is sufficiently large to characterize each resonator's complex resonance circle. On the DAQ personal computer (PC), each block of data is averaged in time to yield a single mean I and Q data point for each carrier tone. As the LO is stepped, the IQ points trace a trajectory in the complex plane for each resonance. From the IQ-plane trajectory, we can determine the optimal power and frequency at which to drive each individual resonator.

\item  Network analyzer mode. The DAC plays back a white noise signal or any other signal that has equal power in all frequency bins. Because we are interested in the information in all FFT bins in this case and cannot send that information in its entirety to the DAQ PC at the 7.5-kHz FFT cycle rate, the time-domain data are coadded prior to the application of an FFT, and the FFT output is then sent to the DAQ PC. The coaddition preserves the signal but averages down the noise; this is because the phases of the transmitted and received signals do not change from one LUT playback cycle to another, but the phase of the noise does vary and is therefore integrated down. The network analyzer mode allows us to perform the following tasks:

\begin{enumerate}

\item Rapid identification of approximate resonator locations.
\item Calculation of the cable delay of the readout setup through a comparison of the phase of each frequency bin.
\item Examination of the current ADC dynamic range to adjust the digital attenuation on the IF board.
\item Verification of system connections.
  \end{enumerate}

\end{itemize}

   A network-analyzer frequency scan can also be implemented using the IQ sweep mode; in this case, the full 491.52-MHz bandwidth is separated into 192 subbandwidths of equal length. We use 192 drive tones to simultaneously scan these subbandwidths and then combine the data to obtain the full 491.52-MHz bandwidth.

   Using the OSR network analyzer mode, we can completely avoid the use of other network analyzers, positioning the OSR as a complete solution for detector readout.

\subsection{Noise study of the readout system}

\subsubsection{Resonator output power and HEMT noise-to-carrier ratio}

In the previous section, we discussed the noise requirements of the OSR system. In combination with the resonator power level, we can calculate the required noise-to-carrier ratio for the OSR system. Each resonator has a power of 10 to 30 pW ($-80$ to $-75$~dBm) at the device.

      Therefore, the noise-to-carrier ratio per hertz at the HEMT will be
\begin{equation}
\resizebox{.9\hsize}{!}{$
\left( \frac{N}{C} \right)_f = \frac{{k_B \times T_N}}{{\mathrm{\text{Readout Power}}}} = \frac{{1.38 \times 10^{-23} \times 2}}{{30 \times 10^{-12}}} = 9.2 \times 10^{-13} / \text{Hz} =  - 120 \text{ dB}/\text{Hz}$}
\end{equation}
Before the HEMT, the 144 resonators will each have a power of $-58.4$
to $-53.4$~dBm. Inside the cryostat, the HEMT (gain of $35$~dB; noise
temperature of $2$~K) and coaxial cable (loss of $3$~dB) result in a timestream noise-to-carrier ratio at the HEMT of $-56.7$ to $-51.7$~dB at a noise bandwidth of 245.76 MHz.

\subsubsection{Noise-to-Carrier ratio of the detector}

The noise study is an important part of the OSR. A complete study of the noise is presented in Appendices A and B, but in this section, an important aspect of the noise study related to the OSR system will be discussed: how the $\left(\frac{N}{C}\right)$ ratio specifications should be translated for the interpretation of measurements.

As indicated in the previous section, the $\left(\frac{N}{C}\right)_f$ ratio leaving the HEMT is $-120.63$~dB/Hz. We refer to this quantity as the frequency--space ${\left( {\frac{N}{C}} \right)}$ ratio: ${\left( {\frac{N}{C}} \right)_f}$ in units of decibels per hertz. We provide a short calculation here to explain the meaning of this value in terms of a power-spectrum measurement.

\paragraph{Single detector.}

     A value of ${\left( {\frac{N}{C}} \right)_f}$ = 1 x $10^{-12}$/Hz implies that the timestream noise-to-carrier ratio will be

\begin{equation}
{\left( {\frac{N}{C}} \right)_t} = {\left( {\frac{N}{C}} \right)_f} \times \text{Nyquist Bandwidth}
\end{equation}
where the Nyquist bandwidth is 245.76 MHz in our case \footnote{Equation 2.2 only holds for $N/C$ in linear units and not in decibels.}.

        Now, if we perform a measurement using a spectrum analyzer, we cannot directly measure ${\left( {\frac{N}{C}} \right)_f}$. This is because the spectrum analyzer treats the carrier and noise in the same manner, either providing the Fourier coefficients of the power or the Fourier spectral density of the power for both, as shown in the following equation:
\begin{equation}
\resizebox{.9\hsize}{!}{$
{\left( {\frac{N}{C}} \right)_{PSD}} = \frac{{PS{D_{noise}}(f)}}{{PS{D_{signal}}(f)}} = \frac{{PS{D_{noise}}(f)}}{{\text{Signal Power}(f)/\Delta f}} = {{\left( {\frac{N}{C}} \right)_{f}}} \times \Delta f$}
\end{equation}
where $\Delta f$ is the FFT bin width
\begin{equation}
\Delta f =\frac{{\text{Nyquist Bandwidth}}}{{\text{Number of PSD Bins}}}
\end{equation}
and we have
\begin{equation}
\text{Noise Power} = {PS{D_{noise}}(f)} \times \text{Nyquist Bandwidth}
\end{equation}

combine the Equation 2.3 and 2.5, we have
\begin{equation}
\resizebox{.9\hsize}{!}{$
{\left( {\frac{N}{C}} \right)_f} = \frac{{PS{D_{noise}}(f)}}{{\text{Signal Power}(f)}} = \frac{{\text{Noise Power}}}{{\text{Signal Power} \times \text{Nyquist Bandwidth}}}$}
\end{equation}

and combine the Equation 2.4 and 2.6, we have

\begin{equation}
\resizebox{.9\hsize}{!}{$
{{\left( {\frac{N}{C}} \right)_{PSD}} = 
{{\left( {\frac{N}{C}} \right)_{f}}} \times \Delta f
= \frac{{\text{Noise Power}}}{{\text{Signal Power} \times \text{Number of PSD Bins}}}}$}
\end{equation}

Here, $PS{D_{noise}}$ and $PS{D_{signal}}$ are the power spectrum densities of the noise and signal, respectively; the former is determined by the HEMT noise temperature, and the latter is determined by the KID readout power and the frequency bin width. Therefore, the measurement we acquire using a spectrum analyzer depends on the frequency-bin width $\Delta f$ that is used. In our case, $\Delta f$ = 7.5 kHz, which corresponds to a 491.52-MHz sampling rate and a $2^{16}$-point FFT. If we use a wider frequency-bin width, we will integrate the HEMT noise over a larger bandwidth without changing the signal power and will thus obtain a larger value of $\left( {\frac{N}{C}} \right)_{PSD}$ for the same $\left( {\frac{N}{C}} \right)_{f}$.

\paragraph{Multiple detectors.}

For multiple tone readouts, let $N_c$ be the number of detectors. The signal power increases linearly with $N_c$, but the number of detectors does not affect the HEMT noise power. Therefore,

\begin{equation}
{\left( {\frac{N}{C}} \right)_{f N_c}} = \frac{{\left( {\frac{N}{C}} \right)_f}}{{{N_c}}}
\end{equation}
The noise-to-carrier ratios $\left( {\frac{N}{C}} \right)$ were used in noise studies of both digital and analog systems for the OSR. Three different units are used throughout this paper. In Table 5 below, the definitions of $\left( {\frac{N}{C}} \right)$ are summarized in these different units, namely, time-domain, frequency-domain, and spectrum-analyzer units, and the conversions among the different units are presented.

\begin{table*}
\caption{Interpretation of various noise-to-carrier ratios.}
\begin{tabular}{|c|c|c|}
\hline \multicolumn{3}{|c|}{Interpretation of Various Noise-to-Carrier Ratios}\\
\cline{1-3}
Time-Domain ${\frac{N}{C}}$ & ${\left( {\frac{N}{C}} \right)_t} = \frac{{\text{Noise Power}}}{{\text{Signal Power}}}$ & dB\\
\hline
Frequency-Domain ${\frac{N}{C}}$ & ${\left( {\frac{N}{C}} \right)_f} = \frac{{\text{Noise Power}}}{{\text{Signal Power} \times \text{Nyquist Bandwidth}}}$ & dB/Hz\\
\hline
Spectrum-Analyzer (PSD) ${\frac{N}{C}}$ & ${{\left( {\frac{N}{C}} \right)_{PSD}} = \frac{{\text{Noise Power}}}{{\text{Signal Power} \times \text{Number of PSD Bins}}}}$ & dB\\
\hline
Conversion Among Ratios & ${\left( {\frac{N}{C}} \right)_f} \times {\rm{\text{Nyquist Bandwidth}}} = {\left( {\frac{N}{C}} \right)_t}$&$\Delta f$\\
~&${\left( {\frac{N}{C}} \right)_{PSD}} = {\left( {\frac{N}{C}} \right)_f} \times {\Delta f}$ & ${ =\frac{{\text{Nyquist Bandwidth}}}{{\text{Number of PSD Bins}}}}$\\
\hline \multicolumn{3}{|c|}{$\text{Noise Power} = {PS{D_{noise}}(f)} \times \text{Nyquist Bandwidth}$
}\\
\hline
\end{tabular}
\end{table*}

\subsection{Data acquisition system and telescope operation}

Heretofore, we have discussed both hardware and firmware. In addition to these components, we also developed a software package to fully automate the process of multiboard readout, hardware control, communication with the telescope, and computer-controlled data acquisition. We call this software package the DAQ system, and a block diagram of this system is presented in Fig. 14.

   The DAQ functions are listed in the block diagram surrounding the hardware and firmware block in the figure.

The preparatory functions for use before the start of observation are on the left-hand side, which include the following:
\begin{enumerate}
\item Initialization of the FPGA firmware. This commands the FPGA to stand by and be ready to process a signal. The required IF board control function and LUT programming are run on the FPGA.
\item IF board setup. The DAQ program sets the parameters for the IF boards, including digital attenuations, digital switches, the FPGA clock, and the LO frequency.
\item LO sweep and generation of resonator frequencies. On the basis of current sky and observation conditions, the IQ sweep mode (Section II.D.6) is used to determine new drive-tone frequencies.
\item LUT and bin selection. On the basis of the newly generated drive-tone frequencies, the LUT and bin-selection buffer on the FPGA are reprogrammed. A second IQ sweep is performed using the new carrier frequencies, which allows us to analyze the resonator properties during the noise-removal and analysis processes\footnote{The second IQ sweep is necessary because the transfer function of the system is a function of both the LO and baseband frequencies and not just of the RF frequency; therefore, the transfer function must be remeasured at the same LO and baseband frequencies that will be used during data acquisition.}.
\item Timestamp generation. An accurate timestamp is generated (Section II.D.5.a) and programmed into the FPGA such that every data packet output by the FPGA contains the correct absolute time-of-day information.
\item Readout system-alive diagnostic. To ensure that the readout system is ready to observe, we run a final system-alive diagnostic to verify that the data output by the FPGA are reasonable by checking that we receive the correct number of carrier tones, that the carrier tones are in the correct order, that the timestamp is correct compared to the clock, that the power level is as expected, and that the on and off resonator carriers behave as expected.
\end{enumerate}
After all of the functions on the left-hand side have been completed, we can begin observing. Thee follow-up functions to be executed once we have raw data from the FPGA output are on the right-hand side and are as follows:
\begin{enumerate}
\item Acquisition of CSO pointing information. The CSO generates 100-Hz pointing information, which is concurrently and separately recorded.
\item Interpretation of the output raw data packets. The output raw data must be converted into an easy-to-read format including separating the on and off resonators, linking the data to the resonators, and combining the timestamps with the telescope pointing information.
\item Noise-removal and data-analysis pipelines. Noise removal is an important part of the system because the astronomical signal contains inherent noise (e.g., sky noise), and the readout electronics introduce additional noise into the raw data. This noise must be removed to obtain the desired SNR, as is discussed in detail in Appendices A and B.
\end{enumerate}

  \begin{figure*}
  \begin{center}
  \includegraphics[width=13cm]{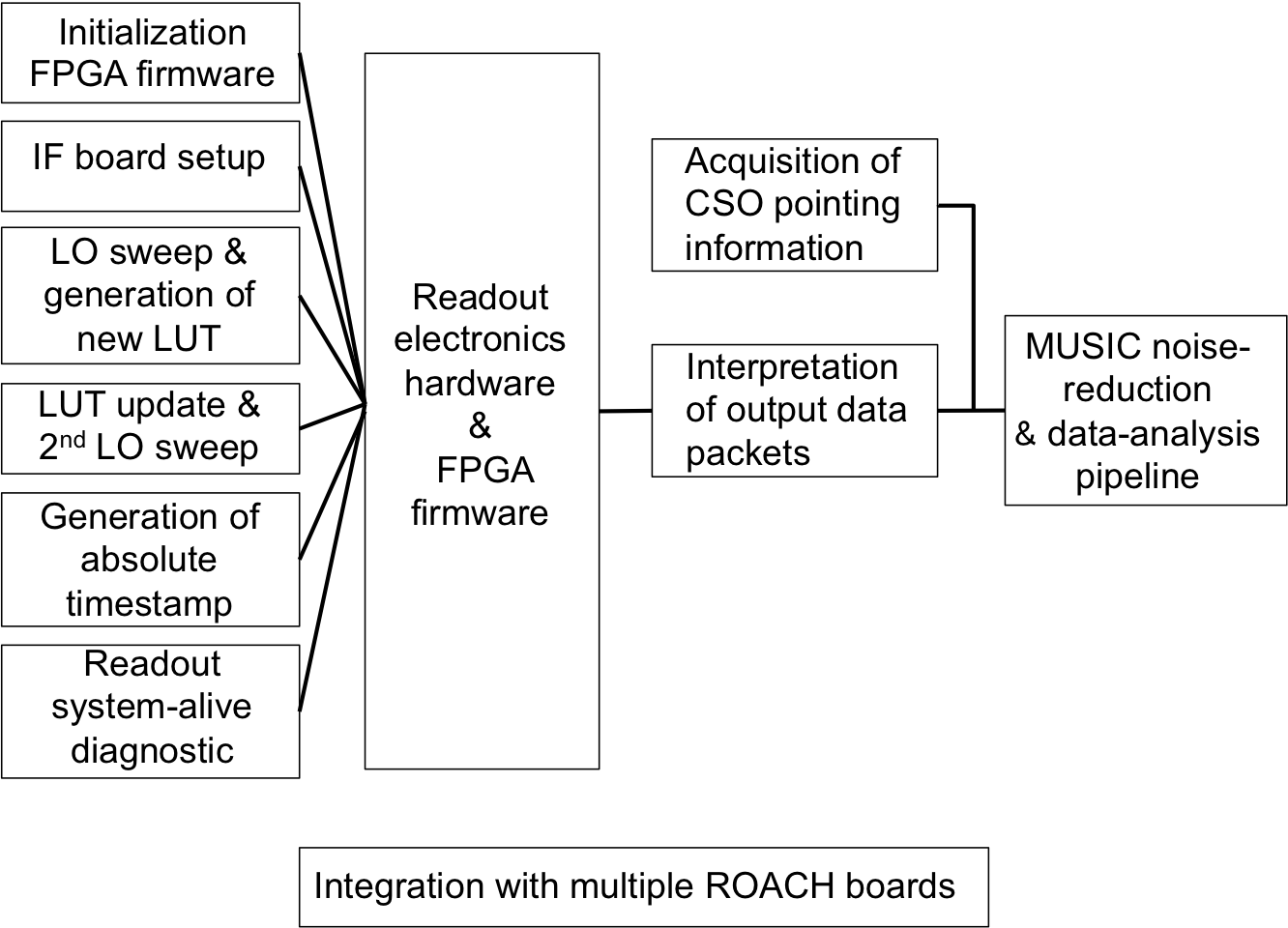}
  \end{center}
  \caption[Block diagram of the DAQ functions and processes.]
  { \label{fig:Block diagram of the DAQ function and process}
Block diagram of the DAQ functions and processes.}
  \end{figure*}

\subsection{Summary of the specifications of the open-source MKID readout system}

The final OSR specifications that we used for the CSO telescope are summarized in Table 6 below.

\begin{table*}
\caption{Summary of the specifications of the open-source MKID readout system.}
\label{tab:fonts}
\begin{tabular}{|l|l|l|}
\hline
\rule[-1ex]{0pt}{3.5ex}  Noise-to-carrier ratio & Less than -123 dB/Hz \\
\hline
\rule[-1ex]{0pt}{3.5ex}  Number of tones & 192   \\
\hline
\rule[-1ex]{0pt}{3.5ex}  Bandwidth & 491 MHz   \\
\hline
\rule[-1ex]{0pt}{3.5ex}  DAC frequency step size & 7500 Hz \\
\hline
\rule[-1ex]{0pt}{3.5ex}  DAC waveform buffer & Continuous; Fast reprogramming  \\
\hline
\rule[-1ex]{0pt}{3.5ex}  Channelizer resolution & 7500 Hz \\
\hline
\rule[-1ex]{0pt}{3.5ex}  Channelizer size & 65536 bins \\
\hline
\rule[-1ex]{0pt}{3.5ex}  Output data format & Complex output with $2 \times 32$ bits  \\
\hline
\rule[-1ex]{0pt}{3.5ex}  Channelizing speed & Real time \\
\hline
\rule[-1ex]{0pt}{3.5ex}  Output data rate & 100 Hz/Channel \\
\hline
\rule[-1ex]{0pt}{3.5ex}  Output protocol & 1-Gbit or 10-Gbit Ethernet; 16 OSR unit switch \\
\hline
\rule[-1ex]{0pt}{3.5ex}  Timestamp & Absolute time of day (precision of up to $1 \times 10^{-8}$ s)  \\
\hline
\rule[-1ex]{0pt}{3.5ex}  OSR synchronization & All electronics components are locked together \\
\hline
\rule[-1ex]{0pt}{3.5ex}  Telescope synchronization & Telescope, electronics and DAQ PC are all locked together \\
\hline
\rule[-1ex]{0pt}{3.5ex}  Operating mode & Many modes available, making the OSR a complete KID readout solution\\
\hline
\rule[-1ex]{0pt}{3.5ex}  Telescope operation  & Fully developed and tested \\
\hline
\rule[-1ex]{0pt}{3.5ex}  Data acquisition  & Fully developed and tested  \\
\hline
\end{tabular}
\end{table*}

\subsection{Discussion of special topics}

In addition to the hardware, firmware, and software, we also made several interesting discoveries regarding the OSR system during the development process.

\subsubsection{Multiple-Tone readout considerations}

Single-tone readout is highly similar to a commercial network analyzer. When ports 1 and 2 are connected to a device, a network analyzer can be used to sweep all frequency bins at a predefined bandwidth and resolution (the $S_{21}$ parameter). Asking the network analyzer to examine only one bin rather than all bins in the bandwidth is similar to reading a single resonator; the multitone readout that we developed is analogous to having 192 network analyzers simultaneously examine 192 bins.

   The readout of many tones from the same transmission line is an extension of the signal-tone readout with some additional criteria:

\begin{enumerate}
\item SNR limitations. If all tones are read using the same DAC and ADC chip, we must ensure that both the individual tone and overall power satisfy the SNR requirements. After the cryogenic amplifier, the ADC is typically the next most important limiting factor for the SNR of the entire readout system.

\item Better adjustment for each tone is required. When we increase the number of tones and begin to utilize all available bandwidth, we must consider the roll-off and ripples across the bandwidth, particularly when we want each drive tone to be optimized at the device, with no loss in power level after propagation along the transmission chain. Thus, there is a need to leave a sufficient margin beyond the theoretical values, which are based on a PAR equal to one.

\item Increased occurrence of spurious tones. The ADC and DAC commonly have an SFDR and IMD that indicate the highest harmonic or spur that the chip will generate. In addition, certain components such as mixers will also introduce spurs. Some spur frequencies are well known, such as the IMD and harmonics. However, other spurs may be randomly distributed across the bandwidth and defined by the SFDR of the chip. To minimize the effects of random spurs, we selected ADC and DAC chips with good SFDRs. For the third-harmonic spurs, because their frequencies are well defined, we can design the carrier frequency bins and program the LUT to ensure that the tones and their harmonics will not overlap.

\item An increased power level may affect the performance of electronic components. For amplifiers and mixers, we must avoid driving the components into their compression ranges. Thus, we should maintain operating limits far away from the 1-dB compression point. For MUSIC, we chose components such that the power levels would be at least $7$~dB away from the compression range. For applications such as MUSIC that require longer-timescale stability with the need to probe signals less than 1 Hz, driving the components excessively hard may introduce additional 1/f noise, as discussed in Section II.H.3.
\end{enumerate}

  \begin{figure}[H]
  \includegraphics[width=8.5cm]{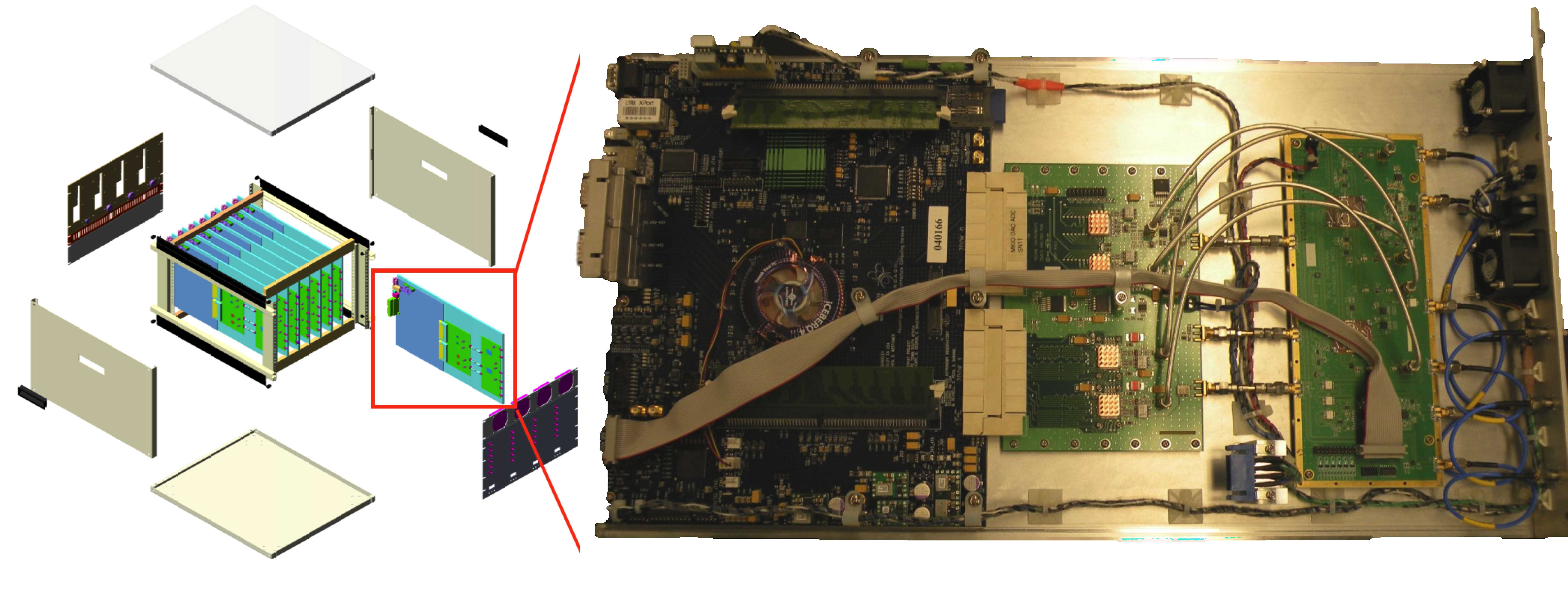}
  \caption[Electronics crate for eight readout units.]
  { \label{fig:Electronic crate for eight readout units}
Electronics crate for eight readout units.}
  \end{figure}

In summary, an OSR for KIDs is analogous to connecting multiple network analyzers on a single transmission line, except that the OSR must read at a predefined time and frequency resolution. Each readout unit (including one ADC/DAC board, one IF board, and one FPGA board) can read out a limited number of resonators (192 carrier tones in the case of MUSIC). We must use a combination of many readout units because there are several thousand resonators in the final telescope instruments. For MUSIC, we integrated eight readout units into a single crate (Fig. 15). This electronic crate not only houses all of the individual readout units together but also provides power through a combined power supply, which significantly facilitates the setup of the readout electronics.


\subsubsection{Discussion of different types of DSP hardware: GPUs and FPGAs}

There are several options for the types of signal processing hardware that can be used for KID readout, including ROACH I, ROACH II, mini-ROACH, or custom-developed boards. There is a tradeoff among the FPGA speed, the readout bandwidth, the FFT length, the number of detectors to read out, and the cost per pixel. We may also need to consider the size of the KID array on the wafer.

   For the MUSIC readout, we implemented all of the required functions using ROACH I. The Zdok connectors are common to ROACH I and II; therefore, the ADC/DAC and IF boards we developed are compatible with both. The yellow blocks\footnote{The input and output simulation blocks in MATLAB Simulink are known as yellow blocks.} for the MKID ADC/DAC would require only a minor modification of the pinout to match a new FPGA on ROACH II. All firmware and software will also operate on ROACH II with an automatic update from the Xilinx compiling tools.

   For KIDs operating at a few gigahertz (as is the case for the MUSIC system), we pack 192 detectors into the 491.52-MHz bandwidth and use a channelizer resolution of 10 kHz. In this case, we may not be able to fully utilize the computing ability of ROACH II, which has a higher-grade FPGA chip. ROACH I is more cost efficient than ROACH II and satisfies our requirements.

   For KIDs operating below 500 MHz, 1000--3000 resonators could fit within this bandwidth. In this case, the readout system will require a larger FFT size (ranging from a few million to a few hundred million bins) on the FPGA. The requirements of a higher FFT size, better frequency resolution, or higher I/O rate indicate that ROACH II may be a more suitable choice.

   Another interesting recent development is that of graphics processing unit (GPU) readout, which combines the functions of an FPGA and a GPU. A GPU is based on a computer operating system and works with a computer motherboard. It offers many advantages including lower cost, ease of installation and programming, and the ability to input and output directly using the computer.

   GPUs are ideal for applications that require signal processing that is more complex than a normal FFT or correlator or nonlinear calculations; however, the speed and input--output (I/O) capability of a GPU are not as good as those of an FPGA. Hence, a combination of an FPGA frontend and a GPU backend is a popular solution. The major calculations that must be performed in a KID application lie in the linear FFT and FIR filter, and these calculations can be largely performed using an FPGA. We have not explored the GPU option further, but the use of a GPU backend would be a suitable choice if more post-FFT analysis becomes necessary in future large KID arrays.

     There has been increasing interest in and effort directed toward the development of an inexpensive customized FPGA/GPU board for KID applications to reduce the cost per pixel. This topic is of interest because the number of KIDs on each telescope will increase in the future to utilize the frequency multiplexing capability of KIDs. ROACH boards are not customized for KID readout, as they contain unnecessary components and features that should be improved upon, and they still have a high cost. This may be a promising direction for future KID arrays.

\subsubsection{System 1/f noise}

Noise in the camera system originates from the readout electronics, HEMT, sky, and MKID device itself. Some of these sources produce uncorrelated noise across different resonators, such as two-level system noise, whereas others produce common-mode noise such as electronics noise.

   We carefully studied the noise performance of the readout electronics, particularly in the low-frequency range. In the 100-Hz audio stream output of the readout system, we observed an increase in noise in both the amplitude and phase directions for frequencies below 10 Hz. Some components of the readout electronics, including the voltage regulator, exhibit such 1/f noise behavior. Clock jitter from the signal synthesizer or voltage-controlled oscillator (VCO) and aperture jitter from the ADC will also appear as low-frequency phase noise.

   To minimize the 1/f noise, we used the frequency-standard locked low-phase-noise option of the VCO on the IF board and used the same VCO as the LO for both up- and down-mixing. We designed a second-generation ADC/DAC board with a low-noise common voltage regulator, clean external DC supply, and better heat sink to reduce low-frequency noise.

  \begin{figure}[H]
  \includegraphics[width=8.5cm]{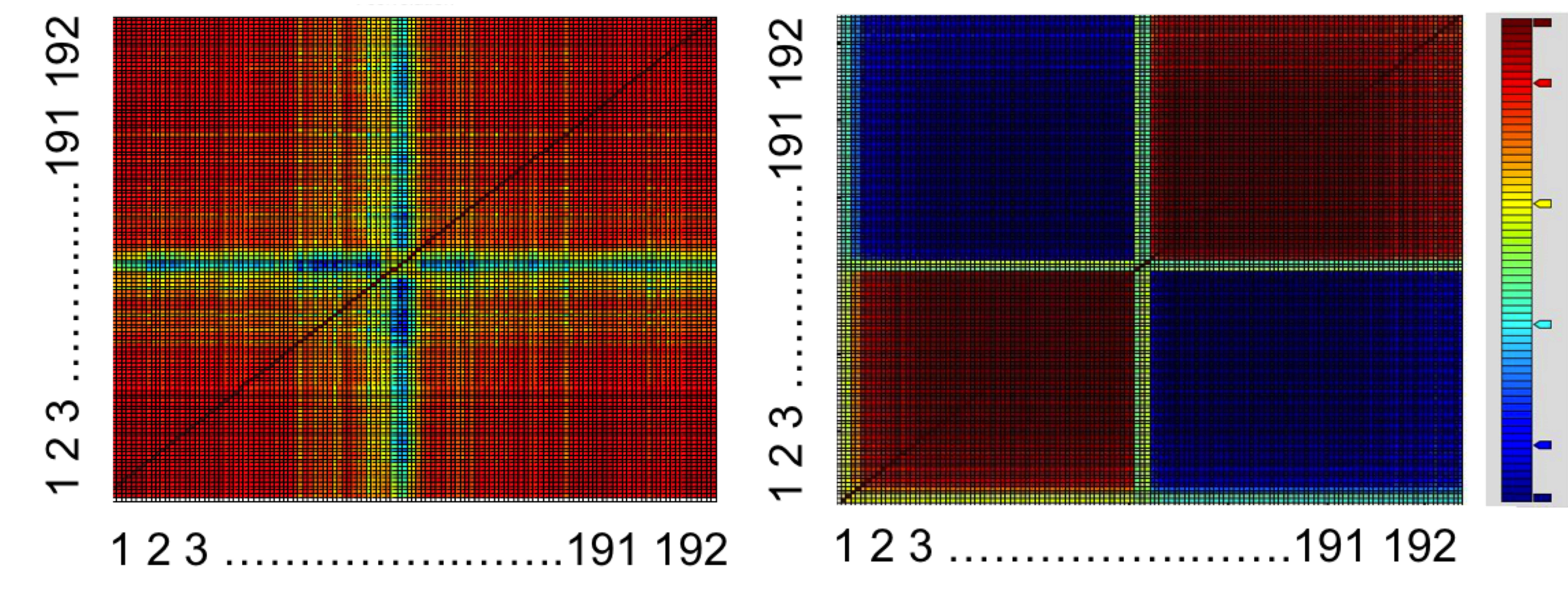}
  \caption[Noise correlation coefficients for the amplitude and phase components.]
  { \label{fig:Noise correlation coefficient of amplitude and phase direction}
Noise correlation coefficients for the amplitude and phase components.}
  \end{figure}

  \begin{figure}[H]
  \includegraphics[width=8.5cm]{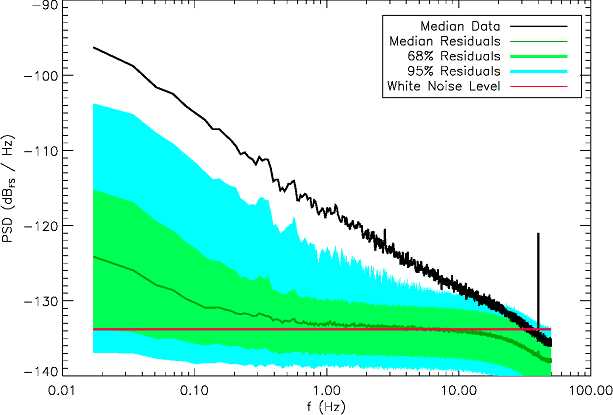}
  \caption[Measured noise performance with noise removal.]
  { \label{fig:Measured noise performance with removal}
Noise power spectral density ($\sqrt{\mathrm{I}^2 +
    \mathrm{Q}^2}$) for the off-resonance tones in laboratory data acquired with
  a stationary dewar. The thick black line represents the raw data,
  the thin green line represents the median post-removal data, and the
  green and cyan bands represent the 68\% and 95\% ranges for the
  post-removal data. The horizontal red line indicates the inferred
  white-noise level, which should be dominated by the HEMT white noise.
  The expectation for the true HEMT noise level is somewhat uncertain
  because of the uncertainty in the system gain from the HEMT to the ADC, but
  it is in the range of $-140$ to $-135$~dBFS/Hz, consistent
  with the measurements. Plot courtesy of Seth Siegel.}
  \end{figure}

      In the OSR setup, we observed a high correlation among all 192 tones in both the amplitude and phase directions (with average correlation coefficients greater than 0.93), as shown in Fig. 16. We can recover a signal that suffers from 1/f readout electronics noise by comparing the on and off resonator tones that were sent out and processed simultaneously, as they experience the same 1/f noise.

   A plot of the noise power spectrum density (PSD) is presented in Fig. 17. In this plot, the red line indicates the measured noise floor of the readout system, which is in agreement with the theoretical calculation of $-135$~dBFS, and the green line indicates the signal noise floor after the removal of 1/f noise, which can reach a theoretical noise floor above 1 Hz.

       When the electronics 1/f noise is combined with the resonator's two-level system noise and the sky noise, the nature of the noise becomes too complex to solve through the simple subtraction of common-mode noise. A detailed and comprehensive study of noise removal performed by our group was integrated into the data-processing pipeline in the DAQ.

\subsubsection{Spurious frequency sources}

Spurious frequencies are detrimental to most KID readouts. The possible causes of spurious frequencies include the following:
\begin{enumerate}
\item Nonlinearity of the IQ mixer. The DAC commonly exhibits good IMD performance; thus, the problem does not originate solely from the DAC output. However, the typical IQ mixer that we use has an input intercept point (IIP) of 10--16~dBm, which means that if we send out a carrier with an amplitude of $-10$~dBm, we should obtain an IMD spur that is approximately $40$~dB lower than the signal at the IQ mixer output. These spurs will appear at the intermodulation products of the input frequencies. The density of intermodulation products is sufficiently high that they may appear to be white noise.

\item DAC quantization. When using MATLAB or Python for sinusoidal calculations, we round the number to the nearest integer, which is a typical source of quantization noise. The theoretical SNR can be calculated to be $6.05N+1.76$, where $N$ is 16 for a 16-bit DAC. This calculation yields an extremely low quantization noise, but a periodic pattern will still be introduced into the LUT timestream, and there will be spurs in the frequency domain that are greater than the theoretical noise level. We have checked this result using a MATLAB simulation and found that this type of spur will be lower than the ADC SNR and will therefore not contribute noise to the system. Although we are still limited by ADC noise, we should be able to visualize this type of spur in the DAC output using the spectrum analyzer.

\item IQ imbalance. When we check the full frequency spectrum of ADC digitized I and Q timestreams (without bin selection and FIR), in addition to the bin with the expected carrier ($BIN_{carrier}$), sometimes a second peak is observed. We define the bin in which this second peak appears at $BIN_{2nd \_ Peak}$. We find this peak exhibits the following behavior: 
\begin{equation}
\text{BIN}_{carrier} + \text{BIN}_{2nd \_ Peak} = \text{total FFT bin number}
\end{equation}
For example, for carrier bin 16382, the 2nd peak appears at bin 114690 for the $2^{17}$ FFT bins.

\begin{figure}[H]
        \centering
        \begin{subfigure}[b]{0.17\textwidth}
                \centering
                \includegraphics[width=\textwidth]{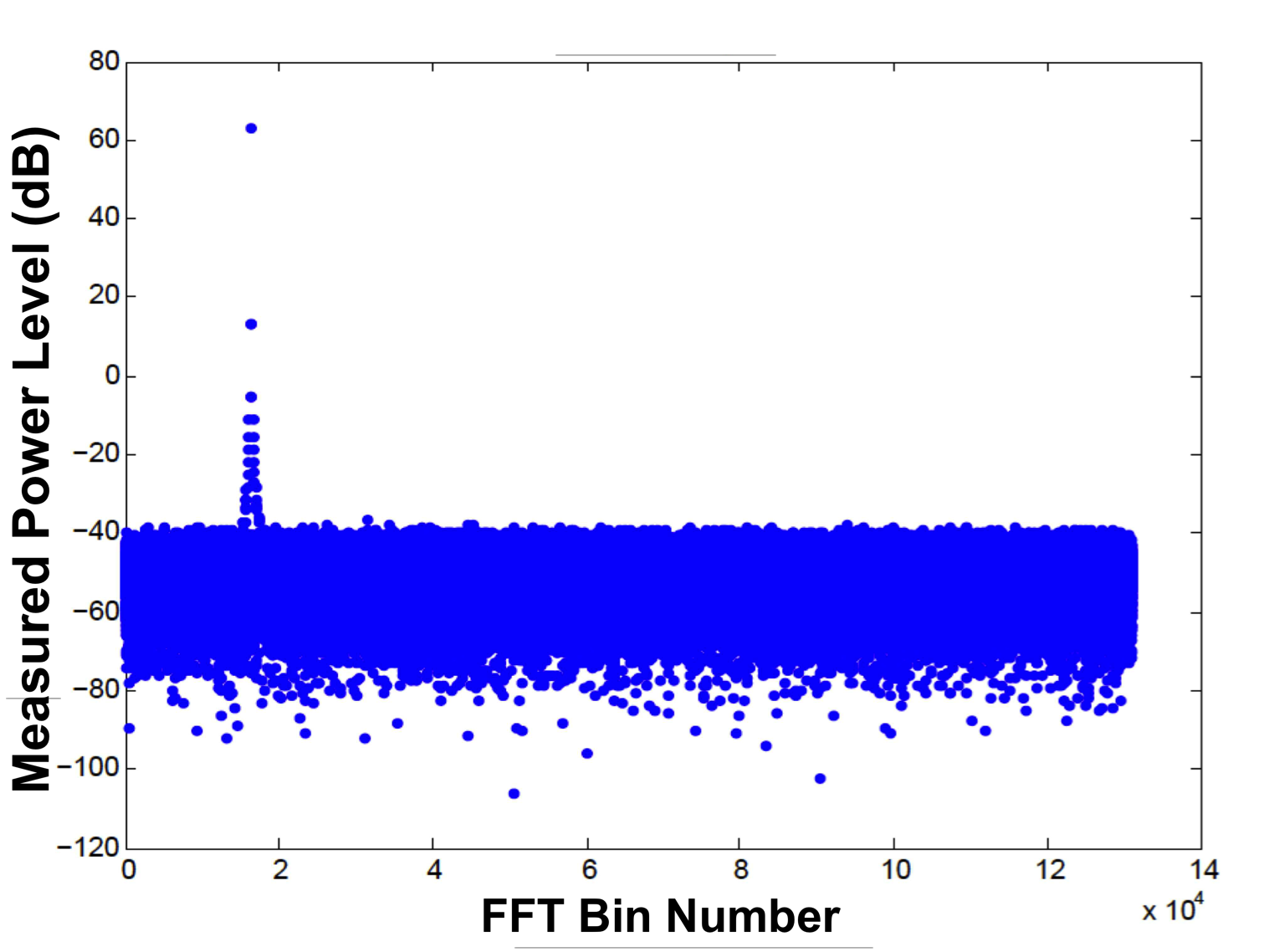}
                \caption{}
                \label{fig:}
        \end{subfigure}%
        ~ 
        \begin{subfigure}[b]{0.17\textwidth}
                \centering
                \includegraphics[width=\textwidth]{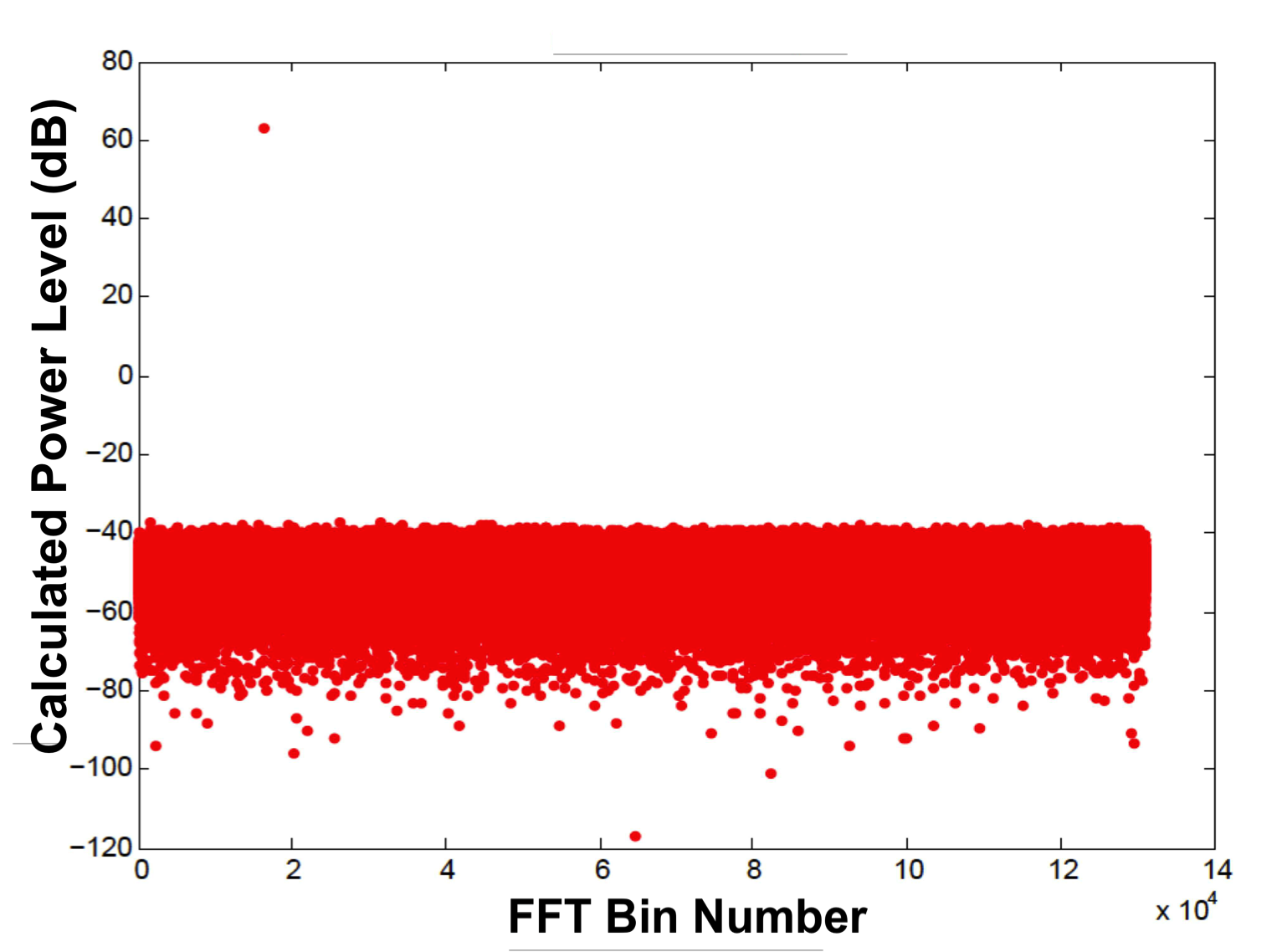}
                \caption{}
                \label{fig:}
        \end{subfigure}
        ~ 
          
        \begin{subfigure}[b]{0.17\textwidth}
                \centering
                \includegraphics[width=\textwidth]{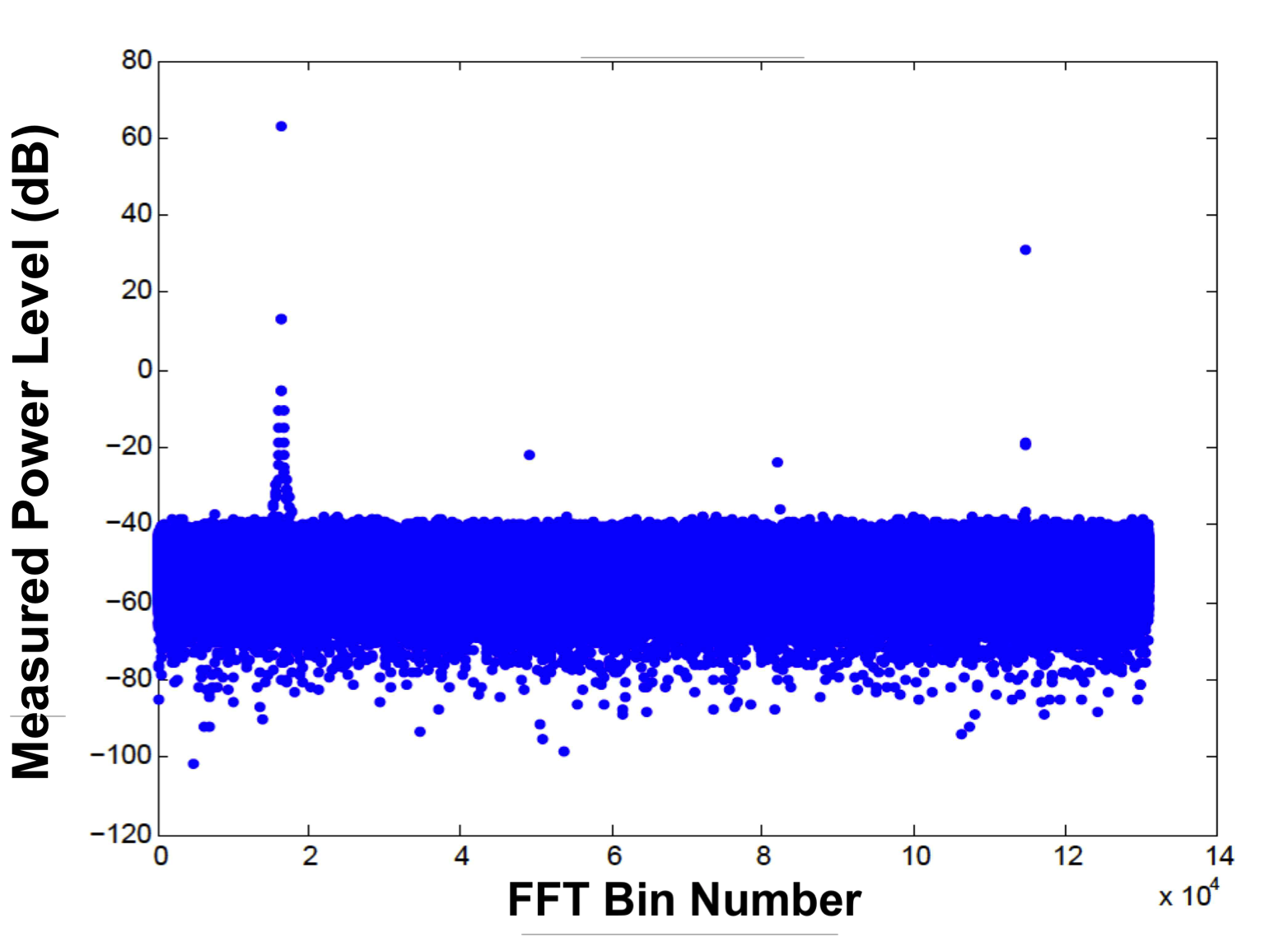}
                \caption{}
                \label{fig:}
        \end{subfigure}%
        ~ 
        \begin{subfigure}[b]{0.17\textwidth}
                \centering
                \includegraphics[width=\textwidth]{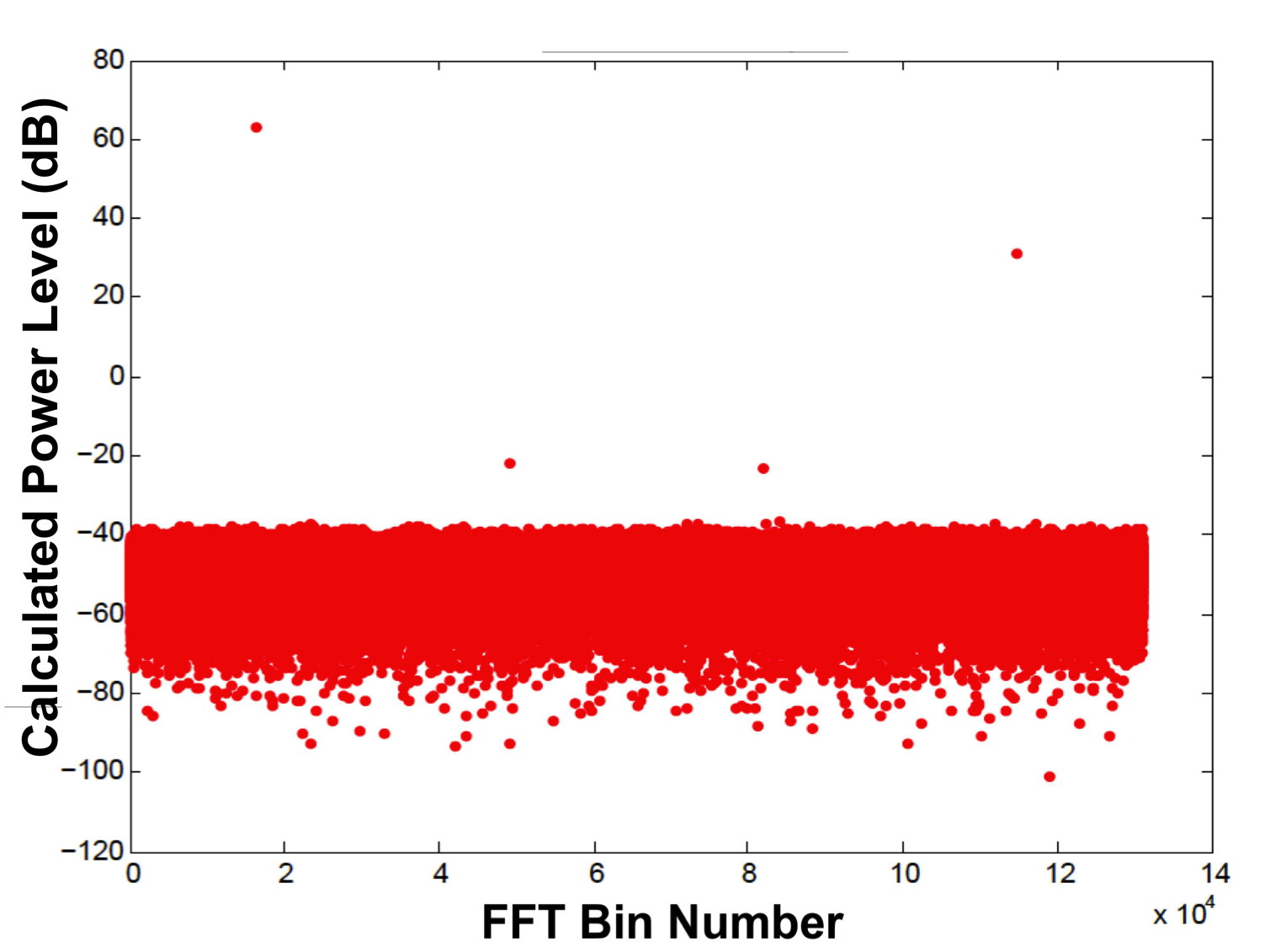}
                \caption{}
                \label{fig:}
        \end{subfigure}
        ~ 

       \begin{subfigure}[b]{0.17\textwidth}
                \centering
                \includegraphics[width=\textwidth]{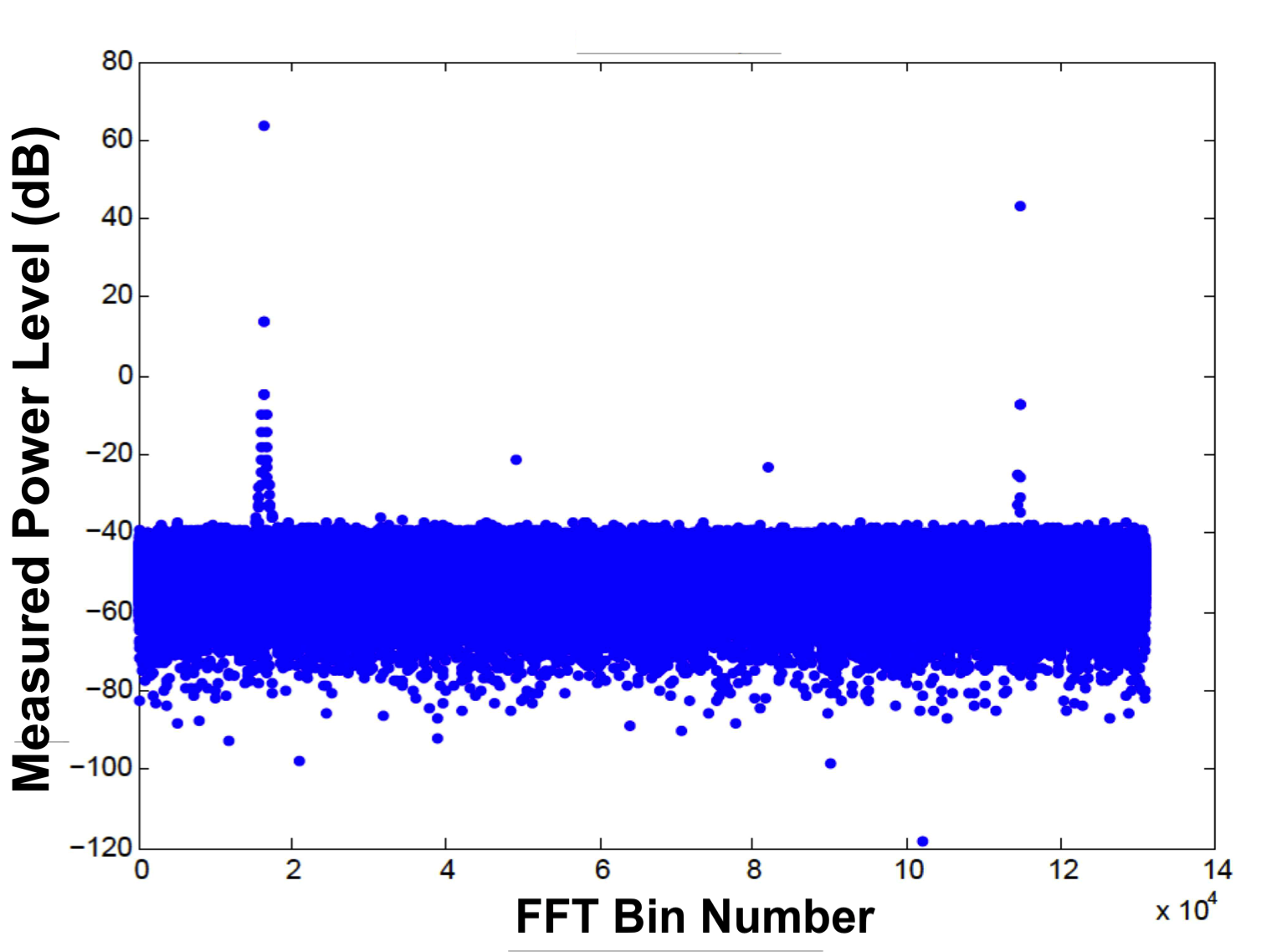}
                \caption{}
                \label{fig:}
        \end{subfigure}%
        ~ 
        \begin{subfigure}[b]{0.17\textwidth}
                \centering
                \includegraphics[width=\textwidth]{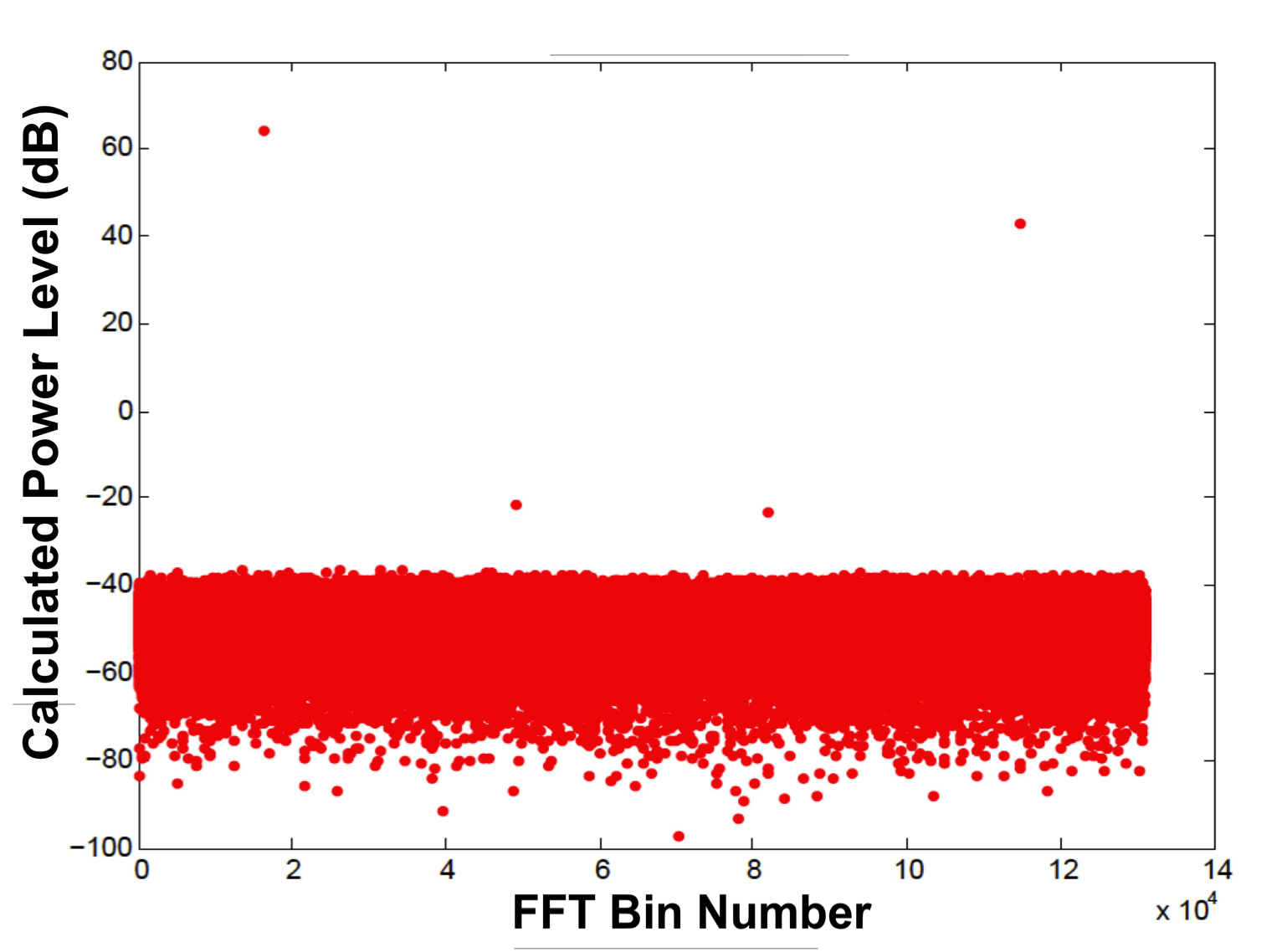}
                \caption{}
                \label{fig:}
        \end{subfigure}

        \caption[Study of the power leakage caused by imbalanced I and Q power levels.]
        {\label{fig:Study of power leakage caused by imbalanced I and Q power levels.}Study of the power leakage caused by unbalanced I and Q power levels. The x axis represents the $2^{17}$ frequency bins; the y axis represents the power level relative to a full-scale of $100$~dBm. (a) and (b) $Amp_I = Amp_Q$, (c) and (d) $Amp_I = 1.05 \times Amp_Q$, and (e) and (f) $Amp_I = 1.2 \times Amp_Q$. (a), (c), and (e) The blue plots represent FPGA complex FFT output of ADC digitized timestreams of DAC/ADC loopback: (FFT(Complex(Digitized\_I, Digitized\_Q))). We store the full FPGA complex FFT output without bin selection and FIR. The amplitude difference between I and Q is achieved by programming the DAC with different LUTs; (b), (c), and (f) the red plots represent the simulation obtained using MATLAB.}
\end{figure}

This result can be explained by the slight difference in the digitized
data between the ADC I and Q inputs. This IQ imbalance can originate from
either the amplitude or phase mismatch between the I and Q signals. For the discussion presented below, take the amplitude imbalance as an example, and let $Amp_I$ and $Amp_Q$ be the amplitudes of the digitized data for the in-phase and quadrature-phase inputs, respectively. For the cases of $Amp_I = Amp_Q$, $Amp_I = 1.05 \times Amp_Q$, and $Amp_I = 1.2 \times Amp_Q$, we studied the effect of this IQ imbalance, as shown in Fig. 18.

From Fig. 18, we can conclude that the measured power leakage due to IQ imbalance agrees with the simulation. The spurious frequency caused by IQ imbalance is well defined in frequency, and can be avoided by carefully picking the carrier tones that do not fall on the leaked frequencies.

\item In addition to the issues mentioned above, other factors (e.g., the delay between the up and down IQ mixers, the IQ mixer DC offset) may also introduce spurs that should be considered in combination with the IIP level; however, these spurious frequencies are more random.
\end{enumerate}

\subsubsection{SNR table}

Fig. 19 presents a table that summarizes the parameters that propagate through the IF board, including the detailed signal power level, noise level, IMD level, accumulated SNR, equivalent noise temperature, and component specifications (model number; gain, loss, and noise figures).

   The table in Fig. 19 was adapted from one provided by an engineer from Omnisys, a vendor that provided an earlier version of the readout system, prior to our OSR development. Under the assumption that the signal power, as the output of the cryogenic HEMT amplifier, is between $-77.5$ and $-55.9$~dBm (for the cases corresponding to 1 and 144 resonators, respectively), we used a 2-K HEMT noise temperature to define the noise level and a 245.76-MHz bandwidth to define the SNR bandwidth as the initial values for the calculation of this SNR table. The remainder of the signal chain could then be designed to inflict minimal SNR degradation.

Consider the first LNA on the IF board (model number HMC753) as an
example; the table indicates that it has a gain of $17$~dB, a noise
figure of $1.5$~dB, and a third-order intercept point output of
$30$~dBm. The table demonstrates that the first-stage LNA will degrade
the SNR by $0.29$~dB for both the 144-resonator case (SNR of
$56.01$~dB) and the one-resonator case (SNR of $34.41$~dB) relative to
the previous stage (the splitter\footnote{Each readout unit (which
  includes one ADC board, one IF board, and one FPGA board on the Al
  plate) covers a bandwidth of approximately 500 MHz, and we use two
  readout units to read out the signal from one feedline that contains
  resonator information in two separate 500 MHz bandwidths. The
  splitter was used to separate the signal into two readout units. The
  splitter in the system is shown the upper right plot in Fig. 22}). The corresponding increase in the noise temperature with respect to the HEMT input is from $2.18$ to $2.33$~K.

This table provides not only the power, noise, and intermodulation values at each stage but also some indication of how the accumulated SNR or noise temperature propagates through the receiving chain. We used this table to select the components, arrange the sequence of components, and optimize the overall SNR and IMD on the IF boards. Once the components were set, we tuned the variable attenuators in the chain to fine-tune the performance. For 144 resonators, the overall SNR degradation is approximately $0.7$~dB; thus, we consider the impact of the designed IF board to be sufficiently low.

  \begin{figure}[H]
  \includegraphics[width=8.5cm]{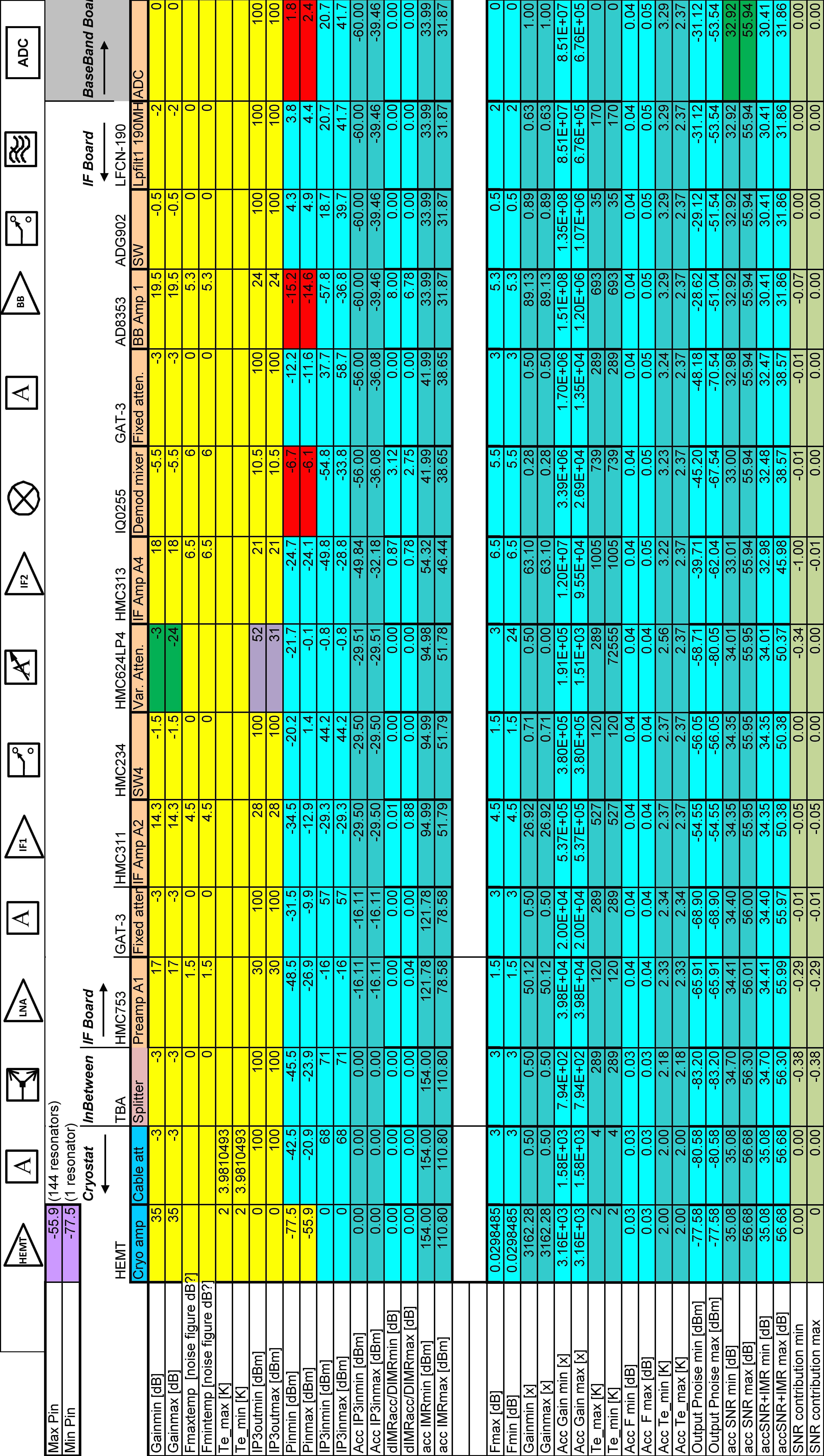}
  \caption[Signal and noise propagation along the analog readout receiving chain.]
  { \label{fig:Signal and noise propagation of Readout analog receiving chain}
Signal and noise propagation along the analog readout receiving chain.}
  \end{figure}

\section{Two engineering runs of MUSIC with the telescope}

In 2010, we tested our MUSIC prototype on the sky at the CSO. At that time, we could read out only 126 resonators; additionally, the system still used many off-the-shelf components including the DC power supply, signal generator, and mixer and used only one ROACH board. As shown in Fig. 20, we even had a computer rotating with the telescope at all times.


         The 2010 telescope engineering run was extremely successful: we tested our MKID wafer, the multicolor imaging, the first version of the FPGA firmware, and most of the DAQ system using the telescope.

  \begin{figure}[H]
  \includegraphics[width=8.5cm]{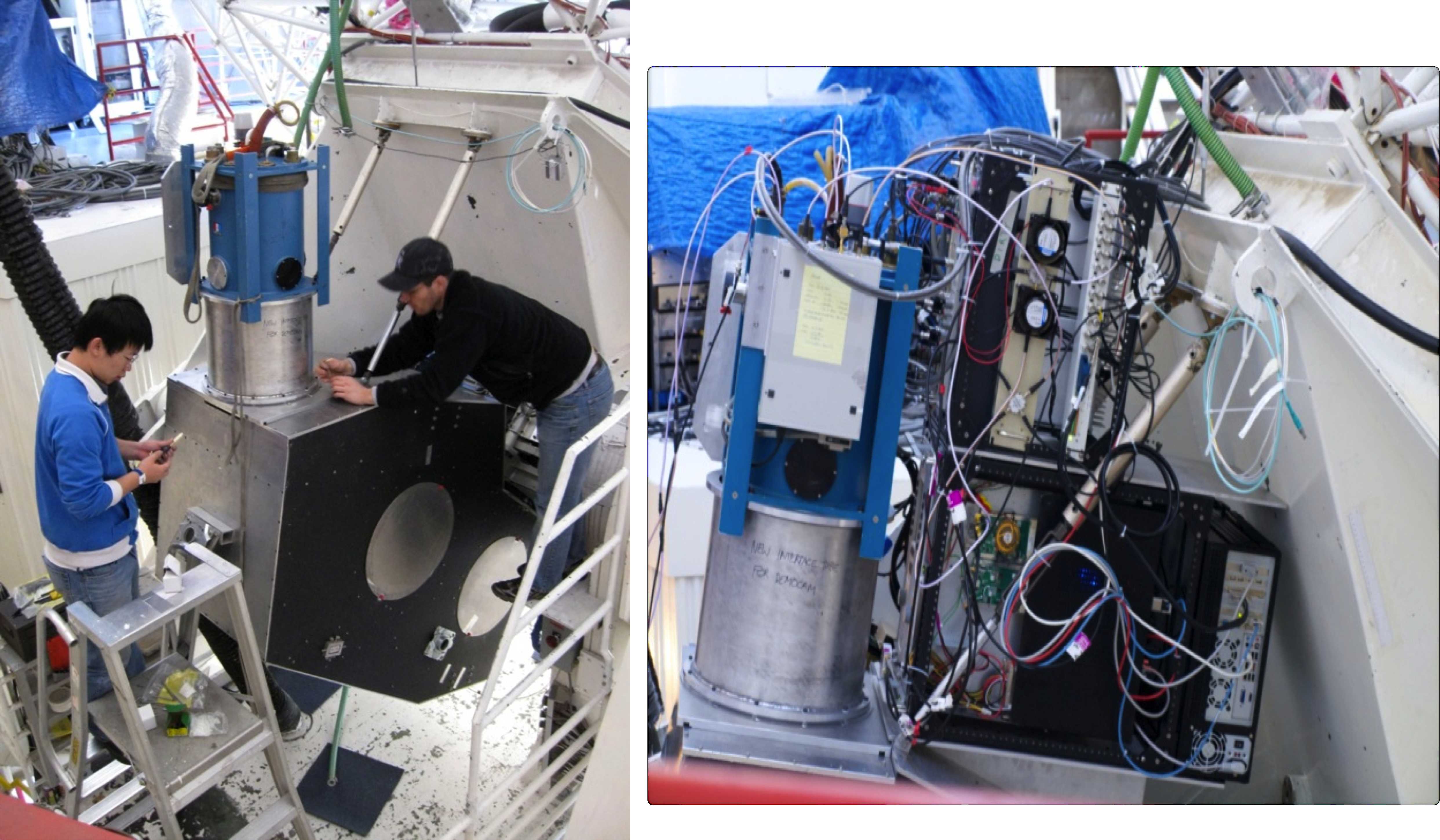}
  \caption[Engineering run at the CSO during Summer 2010: dewar and electronics setup.]
  { \label{fig:Engineering run at CSO during Summer 2010: dewar and electronics setup.}
Engineering run at the CSO during Summer 2010: (left) dewar and optical box installation and (right) electronics and dewar mounted at the CSO.}
  \end{figure}
  \begin{figure}[H]
  \includegraphics[width=8.5cm]{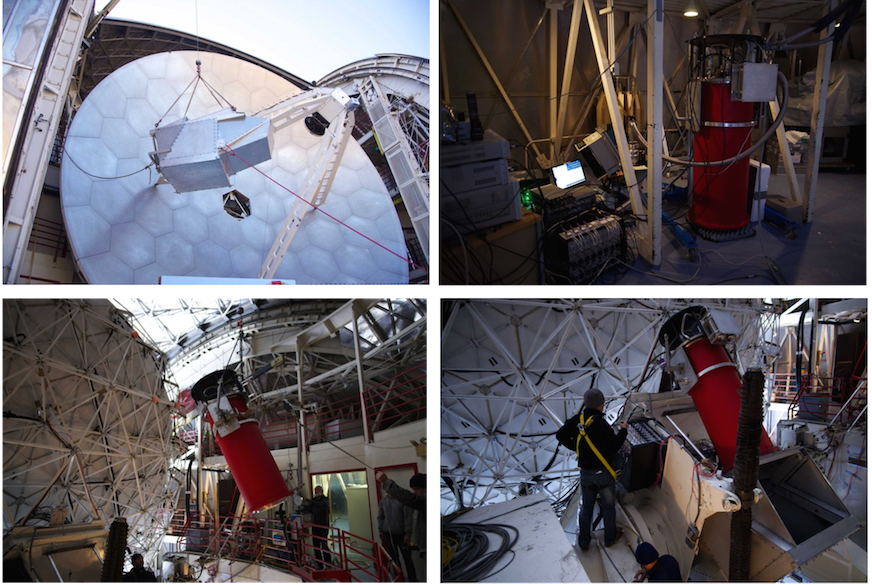}
  \caption[Engineering run at the CSO during Summer 2012.]
  { \label{fig:Engineering run at CSO during Summer 2012.}
Engineering run at the CSO during Summer 2012: (upper left) lifting and installation of the new optical box, (upper right) preparation test of the electronics with the dewar on the third floor of the CSO, (lower left) lifting and installation of the new dewar, and (lower right) installation and mounting of the electronics.}
  \end{figure}


  \begin{figure}[H]
  \includegraphics[width=8.5cm]{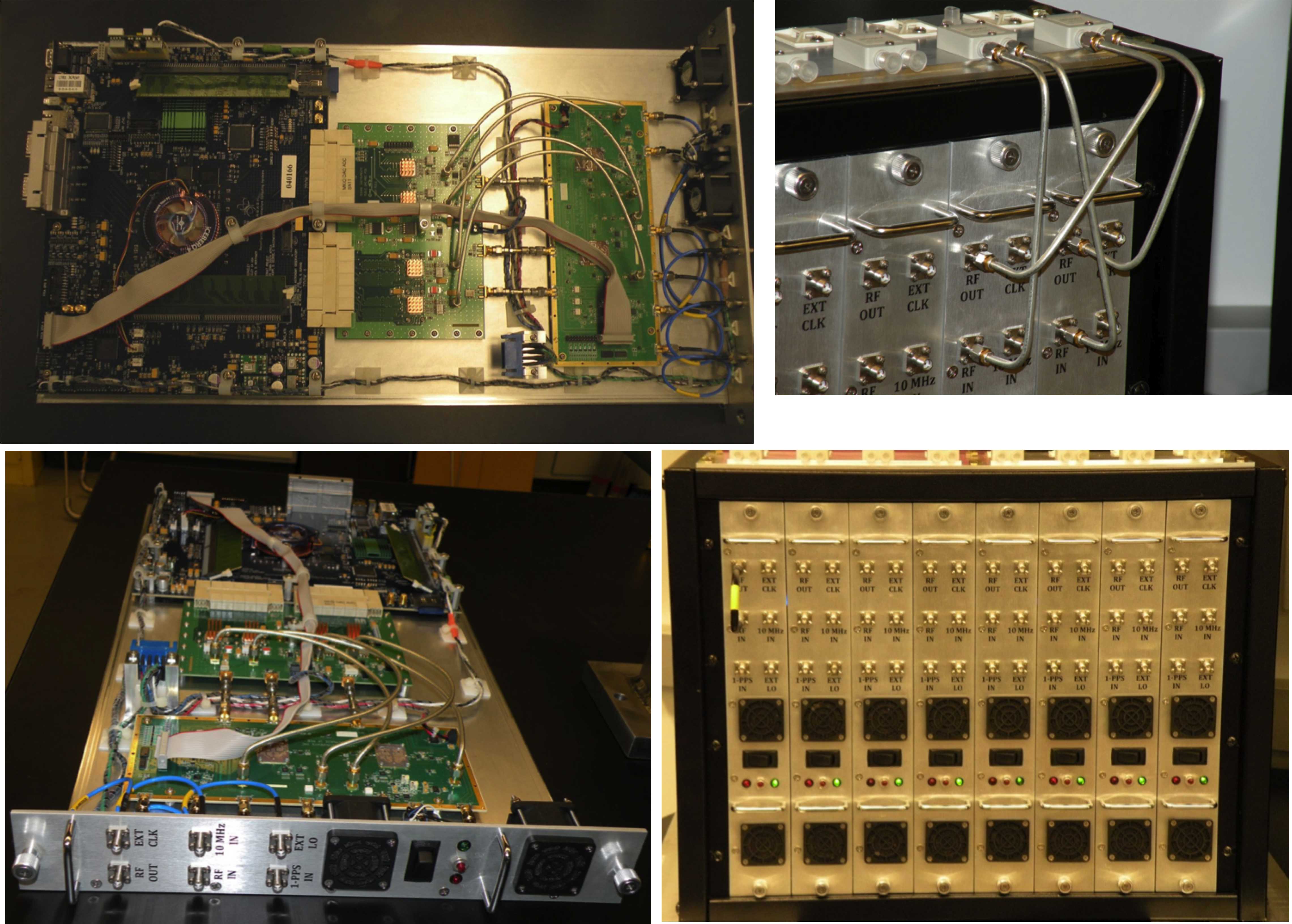}
  \caption[OSR system used in the engineering run at the CSO during Summer 2012.]
  { \label{fig:OSR system used in the engineering run at CSO during Summer 2012.}
OSR system used in the engineering run at the CSO during Summer 2012: (upper left) and (lower left) one unit of OSR electronics, (upper right) the splitter connection on the electronics crate, and (lower right) one complete electronics crate with eight OSR units.}
  \end{figure}

  \begin{figure}[H]
  \includegraphics[width=8.5cm]{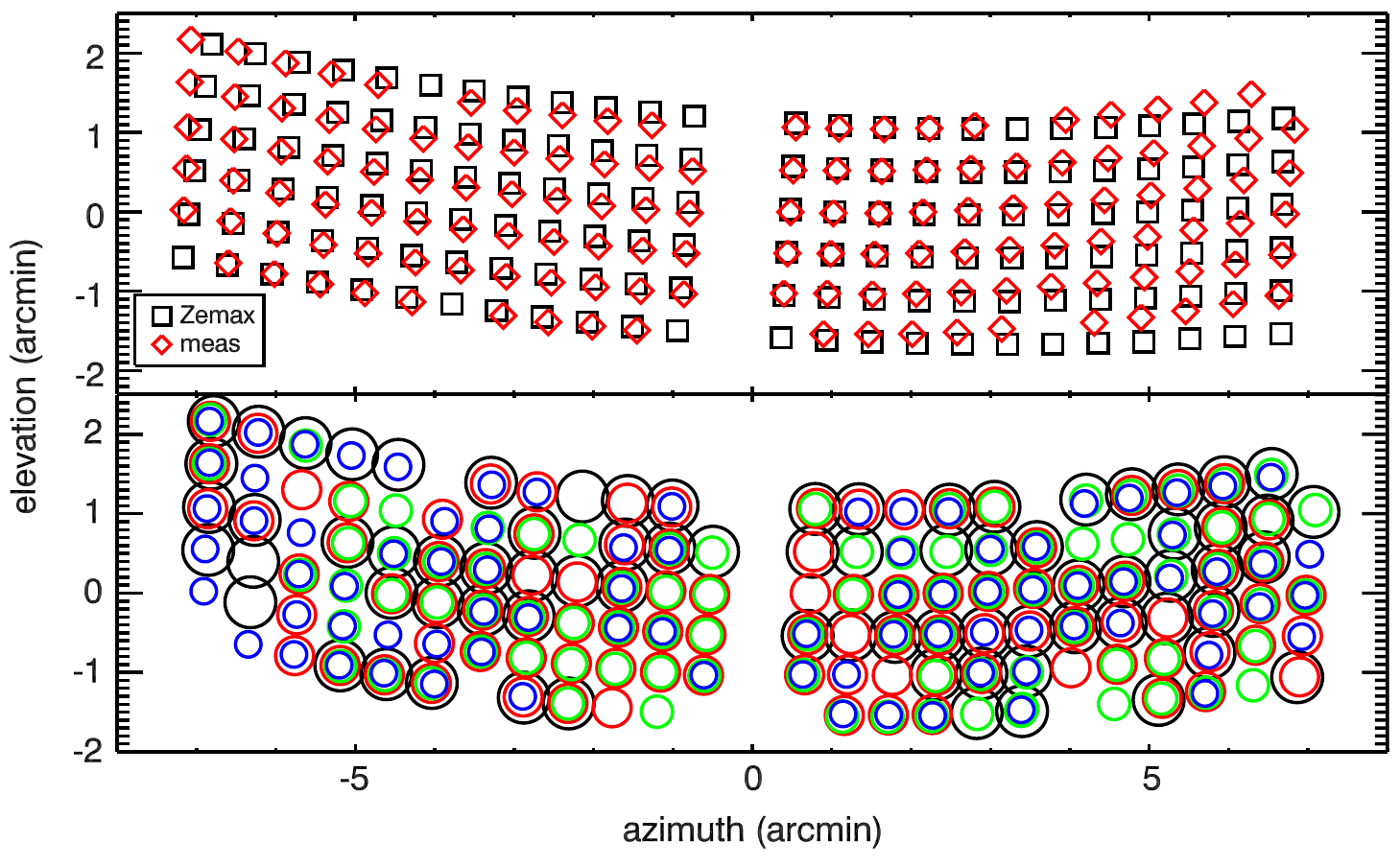}
  \caption[MUSIC mapping of resonators and sky map.]
  { \label{fig:MUSIC mapping of resonator and sky map}
Maps projected onto the sky that were collected by the two subarrays (25\% of the full focal plane) employed during the May 2012 commissioning run based on maps of bright point sources such as Uranus. The top plot compares the observed and expected positions of the beam centroids. The bottom plot shows circles centered on the reconstructed positions with diameters that indicate the reconstructed beam FWHMs and colors that identify the millimeter-wave bands. In the first analysis of these data, all four optical bands were functional in 26 of the 144 total pixels, and three of the four bands were functional in 44 pixels. Plot courtesy of the MUSIC group.}
  \end{figure}

After the first engineering run, we finalized the MUSIC wafer design and completed the final version of the readout system. In 2012, we performed a second engineering run at the CSO. In addition to a new wafer with four-BPF networks and the associated electronics, we implemented a new optical box and cryostat. Fig. 21 shows the installation of the new optical box, new dewar, and new electronics crate, and Fig. 22 shows the final electronics crate that we used. For the OSR system, we implemented two electronics crates, each containing eight OSR units, which allowed us to read out more than 3000 carrier tones. The readout electronics demonstrated their suitability for large-KID-array applications. Fig. 23 shows two subarrays of resonator maps projected onto the sky that were collected during the 2012 run.

  Following the 2012 engineering run, we attempted to improve, complete, and better understand the MUSIC instrument and prepare it to serve as a permanent instrument at the CSO. Since 2013, MUSIC has provided various astronomy groups with submillimeter observations and has started to become widely used by the public. Our group will continue to develop new instruments for submillimeter and millimeter detection. The work that we performed in developing MUSIC at the CSO serves as an excellent and valuable guide for the technological challenges of future submillimeter astronomy.

\begin{acknowledgments}

The projects in this paper were supported by NSF grant AST-0705157 provided to the University of Colorado, a NASA grant provided to the California Institute of Technology, the Gordon and Betty Moore Foundation, and the JPL Research and Technology Development Fund.

\end{acknowledgments}

\bibliography{
AJE_Edited_DJQS5WHB_These_Ran_Duan_2014_July_v13_all_readout}

\end{document}